\documentclass[twocolumn,prd,aps,showpacs,preprintnumbers,nofootinbib,eqsecnum]{revtex4}

\usepackage{amsmath,amsfonts,amssymb}
\usepackage{wrapfig}           
\usepackage{graphicx}          
\usepackage{color}             
\usepackage{dcolumn}           
\usepackage{psfrag}
\usepackage{pstricks,pst-plot} 

\usepackage{t1enc}
\usepackage[english]{babel}   
\usepackage[latin1]{inputenc}

\allowdisplaybreaks

\newcommand{\be}{\begin{equation}}
\newcommand{\ee}{\end{equation}}
\newcommand{\bea}{\begin{eqnarray}}
\newcommand{\eea}{\end{eqnarray}}

\newcommand{\ben}{\begin{eqnarray*}}
\newcommand{\een}{\end{eqnarray*}}

\newcommand{\bs}{\begin{subequations}}
\newcommand{\es}{\end{subequations}}

\newcommand{\vek}[1]{\boldsymbol{#1}}

\begin{document}


\title{Post--Newtonian accurate parametric solution to the
dynamics of spinning compact binaries in eccentric orbits:
The leading order spin--orbit interaction}

\author{Christian K\"onigsd\"orffer}
\email{C.Koenigsdoerffer@uni-jena.de}
\affiliation{Theoretisch--Physikalisches Institut,
Friedrich--Schiller--Universit\"at Jena, Max--Wien--Platz 1,
07743 Jena, Germany}

\author{Achamveedu Gopakumar}
\email{A.Gopakumar@uni-jena.de}
\affiliation{Theoretisch--Physikalisches Institut,
Friedrich--Schiller--Universit\"at Jena, Max--Wien--Platz 1,
07743 Jena, Germany}

\date{\today}

\begin{abstract}
We derive Keplerian--type parametrization for the solution of
post--Newtonian (PN) accurate conservative dynamics of spinning compact
binaries moving in eccentric orbits.
The PN accurate dynamics that we consider consists of the 
third post--Newtonian accurate conservative orbital dynamics 
influenced by the leading order spin effects, namely the leading order
spin--orbit interactions. The orbital elements of the representation are 
explicitly given in terms of the conserved orbital energy,
angular momentum and a quantity that
characterizes the leading order spin--orbit interactions
in Arnowitt, Deser, and Misner--type coordinates.
Our parametric solution is applicable in the following two 
distinct cases:
(i) the binary consists of equal mass compact objects,
having two arbitrary spins, and
(ii) the binary consists of compact objects of arbitrary mass,
where only one of them is spinning with an arbitrary spin.
As an application of our parametrization, we present
gravitational wave polarizations, whose amplitudes are restricted to the
leading quadrupolar order, suitable to describe  
gravitational radiation from spinning compact binaries
moving in eccentric orbits.
The present parametrization will be required to construct 
`ready to use' reference templates for gravitational waves from
spinning compact binaries in inspiralling eccentric orbits.
Our parametric solution for the post--Newtonian
accurate conservative dynamics of spinning compact binaries 
clearly indicates, for the cases considered, the absence
of chaos in these systems.
Finally, we note that our parametrization provides 
the first step in deriving a fully second post--Newtonian
accurate `timing formula', that may be useful for the 
radio observations of relativistic binary pulsars like J0737--3039.
\end{abstract}

\pacs{04.25.Nx, 04.30.Db, 97.60.Jd, 97.60.Lf}

\maketitle

\section{Introduction}
\label{IntroSec}

Inspiralling compact binaries of arbitrary mass ratio, consisting of
black holes and neutron stars, are the most plausible sources of gravitational
radiation for the commissioned ground based and proposed space based
laser interferometric gravitational--wave detectors \cite{IFs}.
The possibility of detecting gravitational radiation from these sources
are based on the following astrophysical and data analysis considerations.
The detailed astrophysical investigations indicate that the inspiral of
compact binaries 
should be observed by gravitational--wave detectors \cite{GLPPS,VK04}.
The temporally evolving gravitational
wave polarizations, 
$h_+(t)$ and $ h_{\times}(t)$,
associated with the inspiralling compact
binaries in quasi--circular orbits are known very precisely \cite{BDFI,ABIQ}.
This allows data analysts to employ the optimal 
method of matched filtering to extract these weak 
gravitational wave signals from the noisy interferometric data 
\cite{BFS91}.

It should be noted that the construction of the `ready to use search
templates', namely above mentioned $h_+(t)$ and $ h_{\times}(t)$,
is possible mainly due to the fact that the orbital dynamics, during
the binary inspiral, is well described by the post--Newtonian (PN)
approximation to general relativity. 
The PN approximation to general relativity 
allows one to express the equations of motion for a compact binary 
as corrections to Newtonian equations of motion
in powers of $(v/c)^2 \sim GM/(c^2 R)$, where $v, M,$ and $R$ are
the characteristic orbital 
velocity, the total mass and the typical orbital separation of the binary.
At present, the orbital dynamics of non--spinning compact binaries are  
explicitly known to 3.5PN order, which gives $(v/c)^7$ corrections
to Newtonian equations of motion.
These lengthy high PN corrections were obtained,
for the first time, in Arnowitt, Deser, and Misner (ADM)--type coordinates
\cite{ADM_S_part1,ADM_S_part2,ADM_S_part3,ADM_S_part4}.
Independently, 
they were later computed in harmonic coordinates
and, as expected, found to be in perfect agreement 
\cite{DD_S_part1,DD_S_part2,DD_S_part3,DD_S_part4,DD_S_part5,DD_S_part6}.
Further, we point out that 
the very recent determination of the highly desirable 
3.5PN (relative) accurate phase
evolution for the gravitational wave polarizations 
was possible, mainly 
because of employing techniques used in the 
computations of 3.5PN accurate equations of motion for the 
dynamics of compact binaries \cite{ADM_S_part3}.

However, the effect of spins and orbital eccentricity on the compact binary
dynamics and the associated 
$h_+(t)$ and $h_{\times}(t)$ are not so accurately determined.
The dynamics of spinning compact binaries are determined  
not only by the orbital equations of motion for these objects, 
but also by the precession equations for 
the spin vectors themselves \cite{Schaefer_grav_effects_2004}.
In fact, for the dynamics of spinning compact binaries
only the leading order contributions to spin--orbit and spin--spin
interactions are well understood \cite{TD_300,Damour2001}.
It is therefore quite customary, in the existing literature,  to consider only
the leading order contributions to spin--orbit and spin--spin 
interactions, while exploring the dynamics of 
spinning compact binaries \cite{CL03} or the 
influence of spins on the phase evolution and amplitude modulations of 
the gravitational waveforms \cite{BCPV04}.
However, note that the equations of motion for spinning compact
binaries, which include first post--Newtonian corrections to the
leading order spin--orbit contributions are also available \cite{TOO04}.
The leading order spins effects on the gravitational wave polarizations, 
mainly for compact binaries in quasi--circular
orbits, were also explored in great detail~\cite{K95}.
As far as the orbital eccentricity is considered, the temporally evolving 
$h_+(t)$ and $h_{\times}(t)$, whose amplitudes are Newtonian accurate
and the orbital evolution is 
exactly 2.5PN accurate, was recently obtained \cite{DGI}.
However, the inputs required to compute the 
crucial secular phase evolution for compact binaries
in inspiralling eccentric orbits are also available to
2PN order beyond the leading quadrupole contributions \cite{GI97,DGI}.
Further, very recently, a Keplerian--type parametrization for the solution of
3PN accurate equations of motion for two non--spinning
compact objects moving in eccentric orbits was obtained \cite{MGS}.
It is therefore desirable to obtain a Keplerian--type parametric solution to
the dynamics of compact binaries, when spin effects are not neglected.

In this paper, we obtain a Keplerian--type parametric solution to the
PN accurate dynamics of spinning 
compact binaries moving in eccentric orbits, 
when the dynamics includes
the leading spin--orbit interaction.
Our parametric solution is applicable in the following two 
distinct cases:
(i) the binary is composed of two compact objects of equal mass and
two arbitrary spins, and
(ii) the two objects have unequal masses but only one object has a spin.
The present parametrization can easily be added to the recently obtained 3PN
accurate generalized quasi--Keplerian parametrization for 
the solution of the 3PN accurate
equations of motion for two non--spinning
compact objects moving in eccentric orbits \cite{MGS}.
This will provide, for the first time, almost analytic,
3PN accurate parametric solution to the 
dynamics of spinning compact binaries moving in eccentric orbits, 
when the spin effects are restricted, for the time being, 
to the leading order spin--orbit contributions \cite{explanation_almost}.
As an application of the parametrization, we present explicit expressions for
the spin--orbit modulated  $h_+(t)$ and $ h_{\times}(t)$, 
whose amplitudes are restricted, for convenience, to the Newtonian order
and the orbital motion is restricted to the above mentioned
conservative PN order.
Our results generalize and improve,
for the two cases considered, a restricted parametrization
for the orbital motion of compact binaries with 
leading order spin--orbit coupling \cite{Wex1995}.

Our parametrization will have the following possible applications.
The construction of `ready to use' search templates,
namely $h_+(t)$ and $h_{\times}(t)$,   
for compact binaries in inspiralling orbits
requires a detailed knowledge of the PN accurate 
orbital dynamics and a PN accurate description for
$h_{ij}^{\rm TT}$, the transverse--traceless (TT) part of the radiation field.
As mentioned earlier, time evolving $h_+(t)$ and $h_{\times}(t)$ for 
non--spinning compact binaries of arbitrary masses
moving in inspiralling eccentric orbits were obtained in Ref.~\cite{DGI}.
The parametrization presented here, along with \cite{MGS}, 
will be required to construct PN accurate `ready to use' reference
templates for spinning compact binaries moving even in 
inspiralling eccentric orbits.

There is an ongoing debate about whether  or not the 
spinning compact binaries, of arbitrary mass ratio, moving in circular
or eccentric orbits under PN accurate dynamics
will exhibit chaotic behavior \cite{CL03}.
The question of chaos was originally motivated
by the observation that the equations describing
the motion of spinning test particle in Schwarzschild spacetime allow
chaotic solutions \cite{SM97}. 
The issue is usually addressed by solving numerically
PN accurate equations describing the dynamics of spinning compact binaries 
to compute certain gauge dependent quantities like fractal
basin boundaries and Lyapunov exponents. 
The semi--analytic parametrization presented here, 
along with the future extensions, 
should be very effective in analyzing the PN accurate 
dynamics of spinning compact binaries 
and to explore if astrophysically realistic binary 
inspiral will be chaotic or not.
Further, the parametrization for the cases considered
implies that the associated PN accurate conservative binary
dynamics can not exhibit chaos.

Finally, we note that our parametrization may be useful to analyze the
high precision radio--wave observations of relativistic binary pulsars like
J0737--3039 \cite{Pulsar_03}. 
These radio--wave observations employ a `timing formula', which gives
the arrival times at the barycenter  
of the solar system for the electromagnetic pulses emitted by a binary pulsar.
The `timing formula' is important for obtaining astrophysical
informations from the compact binary 
as well as to test general relativity in strong field regimes
\cite{DT92,Stairs2003}.
The heavily employed `timing formula', usually incorporates 
1PN accurate orbital motion,
leading order secular effects due to gravitational radiation reaction,
spin--orbit coupling 
as well as 2PN (secular) corrections to 
the advance of periastron \cite{DT92,Timing_S_DD85,Timing_S_DD86,DS88}.
The parametric solution obtained here will be required to obtain a 2PN
accurate `timing formula',
where all spin--orbit and 2PN corrections are fully included.
This implies a `timing formula', which includes not only 2PN order
secular effects, but  
all 2PN order periodic terms, which may leave some 
potentially observable signature \cite{Wex1995,K_04,Kramer_etal_astroph_0409379}.
  
The paper is structured in the following manner. 
In Sec.~\ref{Sec2}, we exhibit and explain, in detail,
the total Hamiltonian describing the conservative dynamics, whose
parametric solution we are seeking in this paper.
In Sec.~\ref{Sec3}, we summarize the Keplerian--type parametrization that 
describes the solution to the 3PN accurate equations of motion for compact 
binaries, in eccentric orbits, when spin effects are ignored.
Sec.~\ref{Sec4} describes the determination of 
a Keplerian--type parametrization for the 
conservative 
dynamics of spinning compact binaries, when spin effects are restricted to the
leading order spin--orbit interaction.
The parametric solutions, presented in Sec.~\ref{Sec3} and
Sec.~\ref{Sec4}, are combined in Sec.~\ref{Sec5}, to obtain
PN accurate Keplerian--type parametrization for the conservative
dynamics of spinning compact binaries moving in eccentric orbits.
As an application to our PN accurate parametric solution, we present 
explicit analytic expressions for the 
gravitational wave polarizations, $h_+(t)$ and $h_{\times}(t)$,  
for spinning compact binaries in eccentric orbits in Sec.~\ref{Sec6}.
The temporal evolution of above $h_+(t)$ and $h_{\times}(t)$, whose amplitudes
are Newtonian accurate, will be governed, for the time being, by the
conservative PN dynamics discussed in this paper.
Finally, in Sec.~\ref{Sec7}, we summarize our results and discuss
its future extensions and applications.

\section{Dynamics of compact binaries with leading order
spin--orbit interaction}  
\label{Sec2}

In this paper, as mentioned earlier, we are interested in the 3PN
accurate conservative dynamics of spinning compact binaries, when spin
effects are, for the time being, restricted to the leading order
spin--orbit interaction. The dynamics is fully specified by a
post--Newtonian accurate Hamiltonian ${\cal H}$, which may be
symbolically written as 
\begin{align}
{\cal H} &
= {\cal H}_{\rm N}
+ {\cal H}_{\rm 1PN}
+ {\cal H}_{\rm 2PN}
+ {\cal H}_{\rm 3PN}
+ {\cal H}_{\rm SO}
\,,
\label{Eq.2.1}
\end{align}
where ${\cal H}_{\rm N}$, ${\cal H}_{\rm 1PN}$, ${\cal H}_{\rm 2PN}$
and ${\cal H}_{\rm 3PN}$
respectively are the Newtonian, first, second and 
third post--Newtonian contributions to the conservative dynamics of
compact binaries, when the spin effects are neglected. The leading
order spin--orbit coupling to the binary dynamics is given
by ${\cal H}_{\rm SO}$. We determine the parametric solution to the
dynamics, given by Eq.~\eqref{Eq.2.1}, in the following way. First, we
obtain a parametric solution to the PN accurate conservative dynamics,
neglecting the effects due to spin--orbit coupling.
We then compute a parametric solution to the conservative dynamics, given by
${\cal H}_{\rm N} + {\cal H}_{\rm SO}$. The second step is consistent in
a post--Newtonian framework as the spin--orbit interactions are restricted to
the leading order. In the final step, we consistently
combine these two parametrizations, to obtain a PN accurate parametric
solution to the conservative compact binary dynamics, as specified by
${\cal H}$.

In the present work, we employ the following 3PN accurate conservative
Hamiltonian, in Arnowitt, Deser, and Misner 
(ADM)--type coordinates \cite{ADM62,ADM_type},
supplemented with a contribution describing the leading order
spin--orbit coupling. We work, as is usual in the literature \cite{MGS},
with the following 3PN accurate conservative reduced Hamiltonian,
$H = {\cal H} / \mu$, where the reduced mass of the binary is given by
$\mu = m_1\,m_2 / M$, $m_1$ and $m_2$ being the individual masses
and $M= m_1 + m_2$ is the total mass.
The 3PN accurate (reduced) Hamiltonian, in ADM--type
coordinates and in the center--of--mass frame,
with the leading order spin--orbit contributions reads 
\begin{widetext}
\begin{align}
\label{H_3}
{H}(\vek{r}, \vek{p}, \vek{S}_{1}, \vek{S}_{2}) &
= {H}_{\rm N}(\vek{r}, \vek{p})
+ {H}_{\rm 1PN}(\vek{r}, \vek{p})
+ {H}_{\rm 2PN}(\vek{r}, \vek{p})
+ {H}_{\rm 3PN}(\vek{r}, \vek{p})
+ {H}_{\rm SO}(\vek{r}, \vek{p}, \vek{S}_{1}, \vek{S}_{2})
\,,
\end{align}
where the Newtonian, post--Newtonian and spin--orbit contributions
are given by 
\begin{subequations}
\label{H_3_Full}
\begin{align}
{H}_{\rm N}(\vek{r}, \vek{p})
&= \frac{\vek{p}^2}{2} - \frac{1}{r}
\,,
\\
{H}_{\rm 1PN}(\vek{r}, \vek{p})
&= \frac{1}{c^2} \left\{
\frac{1}{8} (3\eta-1) \left( \vek{p}^2 \right)^2
- \frac{1}{2} \left[ (3+\eta) {\vek{p}}^2 + \eta(\vek{n} \cdot
\vek{p})^2 \right] \frac{1}{r}
+ \frac{1}{2r^2} \right\}
\,,
\\
{H}_{\rm 2PN}(\vek{r}, \vek{p})
&= \frac{1}{c^4} 
\left\{
\frac{1}{16} \left( 1 - 5\eta + 5 \eta^2 \right)
\left( {\vek{p}}^2 \right)^3
+ \frac{1}{8} \left[
\left( 5 - 20 \eta - 3 \eta^2 \right)
\left({\vek{p}}^2 \right)^2
- 2 \eta^2 (\vek{n} \cdot \vek{p})^2 {\vek{p}}^2 - 3 \eta^2
(\vek{n} \cdot \vek{p})^4 \right] \frac{1}{r}
\right.
\nonumber
\\
&\quad\left.
+ \frac{1}{2} 
\left[ 
(5 + 8 \eta){\vek{p}}^2 + 3 \eta (\vek{n} \cdot \vek{p})^2 
\right] \frac{1}{r^2}
- \frac{1}{4} ( 1 + 3 \eta) \frac{1}{r^3}
\right\}
\,,
\\
{H}_{\rm 3PN}(\vek{r}, \vek{p})
&= \frac{1}{c^6} 
\left(
\frac{1}{128} \left( - 5 + 35 \eta - 70 \eta^2 + 35 \eta^3 \right)
\left( {\vek{p}}^2 \right)^4
+ \frac{1}{16} 
\left[
\left( - 7 + 42 \eta - 53 \eta^2 - 5 \eta^3 \right)
\left( {\vek{p}}^2 \right)^3
\right.
\right.
\nonumber
\\
& \quad \left.
+ ( 2 - 3 \eta) \eta^2 (\vek{n} \cdot \vek{p} )^2
\left( {\vek{p}}^2 \right)^2
+ 3 ( 1 - \eta) \eta^2 (\vek{n} \cdot \vek{p})^4 {\vek{p}}^2
- 5 \eta^3 (\vek{n} \cdot \vek{p})^6 
\right] \frac{1}{r}
\nonumber
\\
& \quad
+ \left[ 
\frac{1}{16} \left( -27+ 136 \eta +109 \eta^2 \right) \left({\vek{p}}^2 \right)^2 
+ \frac{1}{16} ( 17 + 30 \eta) \eta (\vek{n} \cdot \vek{p})^2 {\vek{p}}^2
+ \frac{1}{12}( 5 + 43 \eta) \eta (\vek{n} \cdot \vek{p})^4 
\right] \frac{1}{r^2} 
\nonumber
\\
& \quad
+\left\{ 
\frac{1}{192} [ -600 +\left(3 \pi^2 -1340 \right) \eta -552 \eta^2 ] {\vek{p}}^2
- \frac{1}{64} (340 +3 \pi^2 +112 \eta) \eta (\vek{n} \cdot \vek{p})^2
\right\} \frac{1}{r^3}
\nonumber
\\
& \quad \left.
+ \frac{1}{96} \left[12 +\left( 872 - 63 \pi^2 \right) \eta \right] \frac{1}{r^4}
\right)
\,,
\\
{H}_{\rm SO}(\vek{r}, \vek{p}, \vek{S}_{1}, \vek{S}_{2})
&= \frac{1}{c^2 r^3} (\vek{r} {\times} \vek{p}) \cdot \vek{S}_\text{eff}
\,.
\end{align}
\end{subequations}
\end{widetext}
In the above equations
$\vek{r} = \vek{\cal R}/(G M)$, $r= |\vek{r}|$, $\vek{n}=\vek{r}/r$
and $\vek{p} = \vek{\cal P}/\mu$, where
$\vek{\cal R}$ and $\vek{\cal P}$ are the relative separation vector and
its conjugate momentum vector, respectively.
The finite mass ratio $\eta$ is given by $\eta = \mu /M$.
The spin--orbit coupling involves the effective spin
$\vek{S}_\text{eff}$,
given by
\begin{align}
\vek{S}_\text{eff}
= \delta_{1} \vek{S}_{1} + \delta_{2} \vek{S}_{2}
\,,
\end{align} 
where
\begin{subequations}
\begin{align}
\delta_{1} &
= 2 \eta \left( 1 + \frac{3 m_2}{4 m_1} \right)
= \frac{\eta}{2} + \frac{3}{4} \left(1 - \sqrt{1 - 4 \eta} \right)
\,,
\\
\delta_{2} &
= 2 \eta \left( 1 + \frac{3 m_1}{4 m_2} \right)
= \frac{\eta}{2} + \frac{3}{4} \left(1 + \sqrt{1 - 4 \eta} \right)
\,.
\end{align}
\end{subequations}
The reduced spin vectors $\vek{S}_{1}$ and $\vek{S}_{2}$ are related
to the individual spins $\vek{\cal S}_1$ and $\vek{\cal S}_2$ by 
$\vek{S}_{1} = \vek{\cal S}_1 /(\mu G M)$ and
$\vek{S}_{2} = \vek{\cal S}_2 /(\mu G M)$, respectively.
We recall that $\vek{\cal R}$, $\vek{\cal P}$, $\vek{\cal S}_1$ and
$\vek{\cal S}_2$ are canonical variables, such that the orbital 
variables commute with the spin variables, e.g. see Ref.~\cite{Schaefer_grav_effects_2004,Damour2001}.

The above definition of the effective spin $\vek{S}_\text{eff}$ is
identical to the quantity $\vek{\zeta}$, introduced
in Ref.~\cite{Wex1995}, and related to $\vek{S}_\text{eff} |_{\text{TD}}$,
introduced in Ref.~\cite{Damour2001}, by
$2 \vek{S}_\text{eff} |_{\text{TD}} = G M^2 \vek{S}_\text{eff}$.

The PN corrections, associated with the motion of non--spinning compact
binaries, are compiled using Refs.~\cite{DJS00b,ADM_S_part3}.
The spin--orbit contribution to ${H}$, available in Ref.~\cite{DS88}, employs
the spin--supplementary condition (SSC) of Pryce, Newton and
Wigner~\cite{Pryce_ssc,Newton_Wigner_ssc}.
The SSC, which defines the central world line of each
spinning body, adopted in this paper is given by
\begin{align}
2 {\cal S}_{i0} + \frac{1}{c} {\cal S}_{ij} U^{j} = 0
\quad\quad
(i, j = 1, 2, 3)
\,,
\end{align}
where ${\cal S}_{\mu\nu}$ is the spin--tensor and $U^{\mu}$ the $4$--velocity
of the center of mass of the body. The spin $4$--vector ${\cal S}_{\alpha}$
is defined by
\begin{align}
{\cal S}_{\alpha}
= - \frac{1}{2 c} \varepsilon_{\alpha\beta\mu\nu} U^{\beta} {\cal S}^{\mu\nu}
\quad 
(\alpha, \beta, \mu, \nu = 0, 1, 2, 3)
\,,
\end{align}
where $\varepsilon_{\alpha\beta\mu\nu}$ is the Levi--Civita tensor
with $\varepsilon_{0123} = +1$.
In the rest frame of the body $(U^{\mu} = c \delta^{\mu}_{0} )$
\begin{align}
{\cal S}_{\alpha}
= (0, {\cal S}^{23}, {\cal S}^{31}, {\cal S}^{12})
= (0, \vek{\cal S})
\end{align}
holds. The spins, $\vek{\cal S}_1$ and $\vek{\cal S}_2$, of the
compact objects may be 
given in terms of their moments of inertia and angular velocities of
proper rotations, ${\cal I}_a$ and $\vek{\Omega}_a$ ($a = 1,2$)
respectively, as 
\begin{subequations}
\begin{align}
\vek{\cal S}_1 = {\cal I}_1 \, \vek{\Omega}_1
\,,
\\
\vek{\cal S}_2 = {\cal I}_2 \, \vek{\Omega}_2
\,.
\end{align}
\end{subequations}
Using the above definitions, it is possible to determine at what PN
order the leading order spin--orbit interaction will manifest.
It is easy to see, using Eq.~\eqref{H_3_Full}, that spin effects enter
the dynamics formally at 1PN order. However, for compact objects, the
spin is roughly given by 
${\cal S} \sim m_{\rm co}\, r_{\rm co} \, v^{\rm spin} \sim G \, m_{\rm
co}^2 \, v^{\rm spin} /c^2 $, where $m_{\rm co}, r_{\rm co}$ and
$v^{\rm spin}$ are the typical mass, size and 
rotational velocity of the spinning compact object. 
This implies, for moderate $v^{\rm spin}$ values, that the leading order
spin orbit coupling will enter the compact binary dynamics at ${\cal
O}({1/c^4})$, \textit{i.e.} at 2PN order. It is interesting to note
that if $v^{\rm spin} \sim c$, then the leading order spin--orbit
interaction will numerically manifest at 1.5PN order. However, in that
case the binary dynamics should be dominated by the
currently neglected PN corrections to ${H}_{\rm SO} $. In this
paper, we will, for the sake of clarity and simplicity, assume that
the spin--orbit interaction influence the compact binary dynamics at 2PN order.
We would like to point out that the present derivation of the Keplerian--type 
parametrization for the PN accurate dynamics of spinning compact
binaries does not really care if the spin--orbit coupling manifests at
1.5PN or 2PN order. 

In the next section, we will display and explain the recently obtained
Keplerian--type parametric solution to the 3PN accurate conservative 
orbital motion of two compact objects, of negligible proper rotations,
moving in eccentric orbits~\cite{MGS}. 

\section{Keplerian--type parametrization for the orbital motion of 
compact binaries with negligible proper rotation} 
\label{Sec3}

Using the fully determined 3PN accurate conservative Hamiltonian, in
ADM--type coordinates and in the center--of--mass frame,
available in Refs.~\cite{DJS00b,ADM_S_part3}, very recently a Keplerian--type
parametrization for the orbital motion of compact binaries in eccentric
orbits was derived~\cite{MGS}. 
The derivation of the above parametric solution, usually referred to as the
3PN accurate generalized quasi--Keplerian parametrization,
crucially depends on the following important points.
First, the 3PN accurate conservative Hamiltonian is invariant under 
time translation and spatial rotations, implying 
the existence of 3PN accurate (reduced) energy $E = {H}$
and (reduced) orbital angular momentum $\vek{L} = \vek{r} \times \vek{p}$,
for the binary in the center--of--mass frame.
In particular, the conservation of $\vek{L}$ indicates that
the motion is restricted to a plane, namely the 
orbital plane, and we may introduce polar coordinates such that
$\vek{r} = r ( \cos\varphi, \sin\varphi)$.
Finally, the Hamiltonian equations of motion governing the relative
motion of the compact binary, namely the 3PN accurate expressions for
$\dot{r} = dr/dt$ and $\dot{\varphi}=d\varphi/dt$ --- where t denotes the
coordinate time scaled by $G M$ --- in terms of $E, L = |\vek{L}|$ and $r$,
are polynomials of degree seven in $1/r$.

The third post--Newtonian accurate generalized quasi--Keplerian
parametrization for compact binaries moving in eccentric orbits, in
ADM--type coordinates and in the center--of--mass frame, is given by
\begin{widetext}
\begin{subequations}
\label{e:FinalParamADM}
\begin{align}
\label{Eq.r_u}
r &= a_r \left( 1 - e_r \cos u \right)
\,,
\\
\label{Eq.l_u}
l \equiv n \left ( t - t_0 \right )
&= u - e_t \sin u
+ \left( \frac{g_{4t}}{c^4} + \frac{g_{6t}}{c^6} \right) (v - u)
+ \left( \frac{f_{4t}}{c^4} + \frac{f_{6t}}{c^6} \right ) \sin v
+ \frac{i_{6t}}{c^6} \sin 2 v
+ \frac{h_{6t}}{c^6} \sin 3 v
\,,
\\
\label{Eq.phi_u}
\varphi - \varphi_{0}
&= (1 + k ) \, v
+ \left( \frac{f_{4\varphi}}{c^4} + \frac{f_{6\varphi}}{c^6} \right)
\sin 2 v
+ \left( \frac{g_{4\varphi}}{c^4} + \frac{g_{6\varphi}}{c^6} \right)
\sin 3 v
+ \frac{i_{6\varphi}}{c^6} \sin 4 v
+ \frac{h_{6\varphi}}{c^6} \sin 5 v
\,,
\\
\text{where}\quad
v &= 2 \arctan \left[ \left( \frac{1+e_{\varphi}}{1-e_{\varphi}}
\right)^{1/2} \tan \frac{u}{2} \right]
\,.
\end{align}
\end{subequations}
\end{widetext}
The PN accurate orbital elements $a_r$, $e_r^2$, $n$, $e_t^2$ , $k$,
and $e_{\varphi}^2$, and the PN orbital functions
$g_{4t}$, $g_{6t}$, $f_{4t}$, $f_{6t}$, $i_{6t}$, $h_{6t}$,
$f_{4\varphi}$, $f_{6\varphi}$, $g_{4\varphi}$, $g_{6\varphi}$, $i_{6\varphi}$,
and $h_{6\varphi}$ expressible in terms of $E, L$ and $\eta$ 
are obtainable from Ref.~\cite{MGS} and will be explicitly displayed in
Sec.~\ref{Sec5}, after including the effects due to the leading order
spin--orbit interactions. 

Let us make a few comments about the parametrization.
The radial motion is uniquely parametrized  by Eq.~\eqref{Eq.r_u} in terms 
of some PN accurate semi--major axis $a_r$, radial eccentricity $e_r$ 
and the eccentric anomaly $u$. The angular motion, described by
Eq.~\ref{Eq.phi_u}, is specified by the true anomaly $v$, angular
eccentricity $e_{\varphi}$, advance of the periastron $k$ and some 2PN
and 3PN order orbital functions. 
The explicit time dependence is provided by the 3PN accurate `Kepler
equation', namely 
Eq.~\ref{Eq.l_u}, which connects the mean anomaly $l$,
and hence the coordinate time, to eccentric and true
anomalies $u$ and $v$, respectively.
The PN accurate `Kepler equation' also includes some PN accurate mean
motion $n$, time eccentricity $e_t$ and some 2PN and 3PN order orbital
functions. It should be noted that all these eccentricities, $e_r, e_{\varphi}$
and $ e_t$ are connected to each other by PN accurate expressions in
terms of $E, L$ and $\eta$, given by Eq.~(21) in Ref.~\cite{MGS}.
This implies that it is possible to specify, as in the Newtonian case,
a PN accurate eccentric orbit in terms of, for example,
$a_r$ and $e_r$.
Finally, we recall that the Keplerian--type parametrization for compact
binaries in eccentric orbits at first post--Newtonian order was
obtained in Ref.~\cite{DD85}. Its extension to 2PN order was derived in
Refs.~\cite{DS88,SW93}. 

In the next section, we will improve the above parametrization by including 
the effects due to the leading order spin--orbit interactions.

\section{Quasi--Keplerian parametrization of the conservative dynamics
of a spinning compact binary --- Newtonian motion augmented by
the first--order spin--orbit interaction}
\label{Sec4}

In this section, we incorporate in to the parametrization,
obtained in the previous section, the effects due to the leading order
spin--orbit coupling.
The straightforward way of implementing that is to analyze the orbital
dynamics, which is simply the Newtonian one augmented by the leading order
spin--orbit interaction.
Let us first describe, in the Hamiltonian framework, the binary dynamics
we are interested in.

\subsection{The Newtonian binary dynamics with leading order spin effects}

We investigate the binary dynamics, given by the reduced Hamiltonian 
${H}_\text{NSO}$,  where
${H}_\text{NSO} = {H}_\text{N} +  {H}_\text{SO}$:
\begin{align}
\label{eq:hamilton_NSO}
{H}_\text{NSO} &
=\frac{\vek{p}^2}{2} - \frac{1}{r}
+\frac{1}{c^2 r^3} \vek{L} \cdot \vek{S}_\text{eff}
\,.
\end{align}
The above Hamiltonian prescribes, via the 
Poisson brackets, the following evolution equations for the reduced 
angular momentum vector $\vek{L} = \vek{r} \times \vek{p}$,
and the individual reduced spin vectors $\vek{S}_{1}$ and $\vek{S}_{2}$
\begin{subequations}
\label{eq:d/dt_L_S1_S2_original}
\begin{align}
\label{eq:dLdt_with_Seff}
\frac{d \vek{L}}{dt} &
= \{ \vek{L}, {H}_\text{NSO} \}
= \frac{1}{c^2 r^3} \vek{S}_\text{eff} \times \vek{L}
\,,
\\
\label{eq:dS1dt_with_L}
\frac{d \vek{S}_{1}}{dt} &
= \{ \vek{S}_{1} , {H}_\text{NSO} \}
= \frac{\delta_{1} }{c^2 r^3} \vek{L} \times \vek{S}_{1}
\,,
\\
\label{eq:dS2dt_with_L}
\frac{d \vek{S}_{2}}{dt} &
= \{ \vek{S}_{2} , {H}_\text{NSO} \}
= \frac{\delta_{2} }{c^2 r^3} \vek{L} \times \vek{S}_{2}
\,,
\end{align}
\end{subequations}
where $\{ \ldots , \ldots \}$ denotes the Poisson brackets. 
Note that $d \vek{L} /dt$, given by Eq.~\eqref{eq:dLdt_with_Seff},
gives only the precessional motion of the orbital plane.
We still need to obtain the equations describing the orbital motion.
Before we go onto derive those equations, let us consider the conserved 
quantities associated with ${H}_\text{NSO}$.

The reduced energy $E = {H}_\text{NSO}$ is conserved because of
$\partial_{t} {H}_\text{NSO} = 0$.
Though $\vek{L}$ is not conserved, it is fairly easy to show that 
its magnitude $L = |\vek{L}|$ is conserved:
\begin{align}
\label{eq:L_is_conserved}
\frac{d {L}^{2}}{dt}
&= \frac{d}{dt} (\vek{L} \cdot \vek{L})
= \frac{2}{c^2 r^3} \vek{L} \cdot (\vek{S}_\text{eff} \times \vek{L} )
= 0
\,.
\end{align}
Similarly, below we show that the magnitudes of the spins are also conserved:
\begin{subequations}
\label{eq:S1_S2_magnitude_conserved}
\begin{align}
\frac{d S_1^2}{dt}
&= \frac{d}{dt} (\vek{S}_{1} \cdot \vek{S}_{1})
= \frac{2 \delta_{1} }{c^2 r^3} \vek{S}_{1} \cdot
(\vek{L} \times \vek{S}_{1})
= 0
\,,
\\
\frac{d S_2^2}{dt}
&= \frac{d}{dt} (\vek{S}_{2} \cdot \vek{S}_{2})
= \frac{2 \delta_{2} }{c^2 r^3} \vek{S}_{2} \cdot
(\vek{L} \times \vek{S}_{2})
= 0
\,.
\end{align}
\end{subequations}

Note that Eqs.~\eqref{eq:d/dt_L_S1_S2_original} indicate that 
$\dot{\vek{S}} = - \dot{\vek{L}}$,
where $\vek{S} = \vek{S}_1 + \vek{S}_2 $ is the
total spin reduced by $\mu G M$, the reduced total spin.
We now define the conserved (reduced) total angular momentum vector
\begin{align}
\vek{J} = \vek{L} + \vek{S}
\,.
\end{align}
The above expression times $\mu G M$ gives the
total angular momentum $\vek{\cal J}$, namely
$\vek{\cal J} = \vek{\cal R} \times \vek{\cal P} + \vek{\cal S}$.
Moreover, $\vek{J}$ is conserved both in its magnitude and direction
as $d \vek{J} /dt = 0$ and $dJ/dt = 0$, where $J = |\vek{J}|$.

The fact that $d \vek{J} /dt = 0$ allows us to introduce a reference
orthonormal triad $(\vek{e}_{X},\vek{e}_{Y},\vek{e}_{Z})$, such that 
$\vek{e}_{Z}$ is along the fixed 
vector $\vek{J}$. This gives 
\begin{align}
\label{eq:ansatz_vec_J}
\vek{J} = J \vek{e}_{Z}
\,.
\end{align}
Further, for later use, we choose the line--of--sight unit vector
$\vek{N}$ to lie in the $\vek{e}_{Y}$--$\vek{e}_{Z}$--plane
(see Fig.~\ref{fig:sphere_xyz_ijk}).
This is always possible and reasonable because
of the degree of freedom associated with the choice of $\vek{e}_{X}$ and
$\vek{e}_{Y}$. These unit vectors $\vek{e}_{X}$ and $\vek{e}_{Y}$ span
the invariable plane which is the plane perpendicular to $\vek{J}$.

We now introduce highly advantageous spherical (polar) coordinates
$(r, \theta, \phi)$ and the associated orthonormal triad
$(\vek{n}, \vek{e}_\theta, \vek{e}_\phi)$,
where $\theta$ is the angle between $\vek{e}_{Z}$
and $\vek{n}$ and $\phi$ is the azimuthal angle defined in the invariable
$\vek{e}_{X}$--$\vek{e}_{Y}$--plane
(see Fig.~\ref{fig:sphere_xyz_ijk}),
such that
\begin{align}
\label{eq:vec_n_in_r_theta_phi}
\vek{n} = 
\sin\theta \cos\phi \, \vek{e}_{X} 
+ \sin\theta \sin\phi \, \vek{e}_{Y} 
+ \cos\theta \vek{e}_{Z}
\,.
\end{align}
The relative separation vector $\vek{r}$ and its time derivative 
$\dot{\vek{r}} = d\vek{r}/dr$, written in the orthonormal triad $(\vek{n}, \vek{e}_\theta, \vek{e}_\phi)$ reads
\begin{subequations}
\begin{align}
\label{eq:vec_r_in_r_theta_phi}
\vek{r}
&= r \vek{n}
\,,
\\
\dot{\vek{r}} 
&= \dot{r} \vek{n} 
+ r \dot{\theta} \, \vek{e}_{\theta} 
+ r \sin\theta \dot{\phi} \, \vek{e}_{\phi}
\,,
\end{align}
\end{subequations}
In spherical coordinates, the linear momentum $\vek{p}$ and its magnitude
may be expressed as 
\begin{subequations}
\begin{align}
\vek{p} &=
p_r \vek{n} + p_\theta \vek{e}_\theta + p_\phi \vek{e}_\phi \,,
\\
\vek{p}^2
&= p_r^2 + p_\theta^2 + p_\phi^2
= (\vek{n} \cdot \vek{p})^2 + (\vek{n} \times \vek{p})^2
\nonumber
\\
&= p_r^2 + \frac{L^2}{r^2} \,.
\end{align}
\end{subequations}
With the help of these relations and $E = {H}_\text{NSO}$, the components
of the linear momentum $\vek{p}$ can be expressed using the 
quantities $E$, $L$ and $(\vek{L} \cdot \vek{S}_\text{eff})$ as
\begin{subequations}
\label{eq:comp_of_p_with_cons_quan}
\begin{align}
\label{eq:comp_pr_of_p}
p_r^2 &= 2 E + \frac{2}{r} - \frac{L^2}{r^2}
- \frac{2 ( \vek{L} \cdot \vek{S}_\text{eff}) }{c^2 r^3}
\,,
\\
\label{eq:comp_pphi_of_p}
p_\phi &= \frac{ L_{Z} }{r \sin\theta}
\,,
\\
\label{eq:comp_ptheta_of_p}
p_\theta^2 &= \frac{L^2}{r^2}-p_\phi^2
= \frac{1}{r^2} \left(L^2 - \frac{L_{Z}^2}{\sin^2 \theta} \right)
\,,
\end{align}
\end{subequations}
where $L_{Z} = \vek{e}_{Z} \cdot \vek{L}$. Note that in general 
$L_{Z}$ is not conserved.

The components of $\dot{\vek{r}}$ in the spherical polar coordinates, which also define the orbital equations of motion associated with ${H}_\text{NSO}$,
read
\begin{subequations}
\label{eq:eom_newton_with_so}
\begin{align}
\label{eq:eom_newton_with_so_1}
\dot{r} 
&= \vek{n} \cdot \dot{\vek{r}}
= p_r
\,,
\\
\label{eq:eom_newton_with_so_2}
r \dot{\theta} 
&= \vek{e}_\theta \cdot \dot{\vek{r}}
= p_\theta + \frac{ \vek{e}_\phi \cdot \vek{S}_\text{eff} }{c^2 r^2} 
\,,
\\
\label{eq:eom_newton_with_so_3}
r \sin \theta  \dot{\phi} 
&=  \vek{e}_\phi \cdot \dot{\vek{r}} 
= p_\phi - \frac{ \vek{e}_\theta \cdot \vek{S}_\text{eff} }{c^2 r^2} 
\,.
\end{align}
\end{subequations}
In the above equations, we have used Hamilton's equation
$\dot{\vek{r}} = \partial {H}_\text{NSO}/\partial \vek{p}$, 
$\vek{n} \times \vek{e}_\theta = \vek{e}_\phi$ and
$\vek{n} \times \vek{e}_\phi = - \vek{e}_\theta$.

Summarizing, we note that the dynamics of the binary system, described
by the Hamiltonian ${H}_\text{NSO}$, is also uniquely determined
by the evolution equations given by
Eqs.~\eqref{eq:d/dt_L_S1_S2_original} and
\eqref{eq:eom_newton_with_so}. 
Before trying to find parametric solutions to these equations, let us 
point out several features of the above binary dynamics.

It is easy to write down the precessional equations for 
the (reduced) total spin $\vek{S}$ and the
effective spin $\vek{S}_\text{eff}$ as
\begin{subequations}
\label{eq:d/dt_for_S_and_Seff}
\begin{align}
\label{eq:dSdt_with_L}
\dot{\vek{S}}  
&= \dot{\vek{S}}_{1} + \dot{\vek{S}}_{2}
= \frac{1}{c^2 r^3} \vek{L} \times \vek{S}_\text{eff}
= - \dot{\vek{L}}
\,,
\\
\label{eq:dSeffdt_with_L}
\dot{\vek{S}}_{\text{eff}} 
&= \delta_{1} \dot{\vek{S}}_{1} +\delta_{2} \dot{\vek{S}}_{2}
=\frac{1}{c^2 r^3}  \vek{L} \times ( \delta_{1}^{2} \vek{S}_{1} +
\delta_{2}^{2} \vek{S}_{2} )
\,.
\end{align}
\end{subequations}
It is clear that the motions of $\vek{S}$ and 
$\vek{S}_\text{eff}$ are more complicated than those for
$\vek{S}_1$ and $\vek{S}_2$.
We deduce that the magnitudes of $\vek{S}$ and $\vek{S}_\text{eff}$
are not preserved in general and satisfy following equations
\begin{subequations}
\label{eq:S_and_S_eff_mag_variation}
\begin{align}
\label{eq:S_mag_variation}
\frac{d S^{2}}{dt}
&= - \frac{ 3 \sqrt{ 1 - 4 \eta } }{c^2 r^3}
\vek{L} \cdot (\vek{S}_1 \times \vek{S}_{2})
\,,
\\
\label{eq:Seff_mag_variation}
\frac{d S^2_\text{eff}}{dt}
&= - \frac{ 3 \sqrt{ 1 - 4 \eta } (12 \eta + \eta^2 )}{ 4 c^2 r^3 }
\vek{L} \cdot (\vek{S}_1 \times \vek{S}_{2})
\,.
\end{align}
\end{subequations}

We note that the quantity $(\vek{L} \cdot \vek{S}_\text{eff})$
is exactly preserved since
\begin{align}
\label{eq:L_dot_Seff_conserved}
\frac{d}{dt} (\vek{L} \cdot \vek{S}_\text{eff})
&= \frac{d \vek{L}}{dt} \cdot \vek{S}_\text{eff}
+ \vek{L} \cdot \frac{d \vek{S}_\text{eff}}{dt}
= 0
\,.
\end{align}

Finally, let us consider the evolution of $L_{Z}$.
Using $\vek{L} = J \vek{e}_Z - \vek{S}$ on the r. h. s. of
Eq.~\eqref{eq:dLdt_with_Seff} we obtain
\begin{align}
\frac{d L_Z}{dt}
&= \frac{3 \sqrt{ 1 - 4 \eta } }{2 c^2 r^3}
\vek{e}_{Z} \cdot (\vek{S}_1 \times \vek{S}_2)
\,.
\end{align}
The above equations imply that, in general, $L_{Z}$ is not conserved.
However $L_{Z}$ along with $S$ and $S_\text{eff}$ are conserved for
the following two cases.
In the first instance, case (i), we require the binary to have equal masses,
but with arbitrary spins $\vek{S}_1$ and $\vek{S}_2$
($m_{1} = m_{2} \Leftrightarrow
\eta = 1/4 \Leftrightarrow
\delta_{1} = \delta_{2}$).
In the second instance, case (ii), the binary may have arbitrary masses, but
only \emph{one} of them is spinning
($m_{1} \neq m_{2}$, $\vek{S}_1 \neq 0$ \emph{or} $\vek{S}_2 \neq 0$).

In this paper, when we solve the differential equations,
Eqs.~\eqref{eq:d/dt_L_S1_S2_original} and \eqref{eq:eom_newton_with_so},
we restrict ourselves, for the time being, to the above two cases.
In these cases, the effective spin $\vek{S}_\text{eff}$ and the
reduced total spin $\vek{S}$ are related by
\begin{align}
\label{eq:Seff_eqauls_coupling_constant_S}
\vek{S}_\text{eff}
= \delta_{1} \vek{S}_{1} + \delta_{2} \vek{S}_{2}
= \chi_\text{so} \vek{S}
\,,
\end{align}
where the newly defined mass dependent coupling constant 
$\chi_\text{so}$ is given by
\begin{equation}
\label{eq:coupling_constant_in_cases}
\chi_\text{so} :=
\begin{cases}
\delta_{1} = \delta_{2} = 7/8 
& \text{for (i), the equal--mass--case}
\,,
\\
\delta_{1} \; \text{or} \; \delta_{2}
& \text{for (ii), the single--spin--case}
\,.
\end{cases}
\nonumber
\end{equation}
We would like to emphasize that in the case (i)
the reduced total spin $\vek{S}$ 
is the sum of two arbitrary spins 
$\vek{S}_{1}$ and $\vek{S}_{2}$. However, in the 
case (ii) $\vek{S}$ is either 
$\vek{S}_{1}$ \emph{or} $\vek{S}_{2}$.

These observations allow us to present the precessional equations 
for $\vek{L}$ and $\vek{S}$ in the following compact form for the
special cases as 
\begin{subequations}
\label{eq:L_and_S_precess_about_J}
\begin{align}
\label{eq:L_precess_about_J}
\frac{d \vek{L}}{dt} &= \frac{\chi_\text{so}}{c^2 r^3} \vek{J} \times \vek{L}
\,,
\\
\label{eq:S_precess_about_J}
\frac{d \vek{S}}{dt} &= \frac{\chi_\text{so}}{c^2 r^3} \vek{J} \times \vek{S}
\,,
\end{align}
\end{subequations}
The Eqs.~\eqref{eq:L_and_S_precess_about_J} imply that the
(reduced) orbital angular momentum $\vek{L}$ and the (reduced)
total spin $\vek{S}$, for the two cases considered,
precess about the fixed vector $\vek{J} = J \vek{e}_{Z}$ at the same rate
with an (instantaneous) precession frequency given by
\begin{align}
\label{eq:definition_of_omega_p}
\omega_p = \frac{\chi_\text{so} J}{c^2 r^3}
\,.
\end{align}
Note that $\omega_p$ can only be defined in a \emph{physically} 
meaningful way when spin effects are included in the binary dynamics. 
In the non--spinning limit, the cross products appearing 
on the r. h. s. of Eqs.~\eqref{eq:L_and_S_precess_about_J} vanish,
leaving the (reduced) orbital angular momentum $\vek{L}$ as a constant vector.
It is interesting to observe that in the non--spinning limit, neither 
$\omega_p$ nor $\chi_\text{so}$ go to zero.

In the next subsection, we will find a parametric solution to the
dynamics of spinning compact binaries in eccentric orbits, given by
Eqs.~\eqref{eq:eom_newton_with_so} and \eqref{eq:L_and_S_precess_about_J}.

\subsection{The Keplerian--type parametrization associated with  
the Hamiltonian ${H}_\text{NSO}$}

We start by considering the radial motion, governed by
Eq.~\eqref{eq:eom_newton_with_so_1}, which reads 
\begin{align}
\label{eq:r_dot_square_in_E_L_LdotSeff}
\dot{r}^2
= 2 E + \frac{2}{r} - \frac{L^2}{r^2}
- \frac{2 ( \vek{L} \cdot \vek{S}_\text{eff}) }{c^2 r^3}
\end{align}

We obtain parametric solution to the above equation by
following exactly the same procedure described in detail in Sec.~III of
Ref.~\cite{MGS}. The radial motion is 
described by
\begin{subequations}
\label{r_and_n(t_t0)}
\begin{align}
\label{eq:radial_SO}
r &= a_r (1 - e_r \cos u )
\,,\\
\label{eq:n(t_t_0)_SO}
l \equiv n(t-t_0) &= u - e_t \sin u
\,,
\end{align}
\end{subequations}
where $u$ and $l$ are the eccentric and mean anomalies.
The orbital elements, explicitly given in terms of 
$E$, $L$ and $\vek{L} \cdot \vek{S}_\text{eff}$, are 
\begin{subequations}
\begin{align}
a_r &=-\frac{1}{2 E}\left(1-2 \frac{\vek{L} \cdot \vek{S}_\text{eff}}{L^2}
\frac{E}{c^2}\right)
\,,
\\
e_r^2 &= 1+ 2 E L^2 + 8 (1 + E L^2)
\frac{\vek{L} \cdot \vek{S}_\text{eff}}{L^2} \frac{E}{c^2}
\,,
\\
n &= (- 2 E)^{3/2}
\,,
\\
e_t^2 &= 1+ 2 E L^2 + 4 \frac{\vek{L} \cdot \vek{S}_\text{eff}}{L^2}
\frac{E}{c^2} 
\,.
\end{align}
\end{subequations}
Eq.~\eqref{eq:n(t_t_0)_SO} gives the `Kepler equation', modified 
by the spin--orbit interaction, which connects the eccentric anomaly $u$ to
the coordinate time. The crucial requirements to determine the above 
parametrization are the conservation of $E$, $L$ and 
$\vek{L} \cdot \vek{S}_\text{eff}$. This allows us to treat
$(dr/dt)^2$ as a polynomial in $1/r$ with constant coefficients.

\begin{figure}[ht]
  \centering
  \psfrag{X}{$\vek{e}_X$}
  \psfrag{Y}{$\vek{e}_Y$}
  \psfrag{Z}{$\vek{e}_Z$}
  \psfrag{J}{$\vek{J}= J \vek{e}_Z$}
  \psfrag{N}{$\vek{N}$}
  \psfrag{n}{$\vek{n} = \vek{r} / r$}
  \psfrag{iTheta}{$\vek{i}$}
  \psfrag{j}{$\vek{j}$}
  \psfrag{k}{$\vek{k}=\vek{L}/L$}
  \psfrag{phi1}{$\phi$}
  \psfrag{phi2}{$\varphi$}
  \psfrag{theta1}{$\theta$}
  \psfrag{theta2}{$\Theta$}
  \psfrag{Upsilon}{$\Upsilon$}
  \psfrag{lineofsight}{\small{Line of sight}}
  \psfrag{invariable}{\small{Invariable plane}}
  \psfrag{orbit}{\bf{\small{Orbital plane}}}
  \includegraphics[scale=0.65]{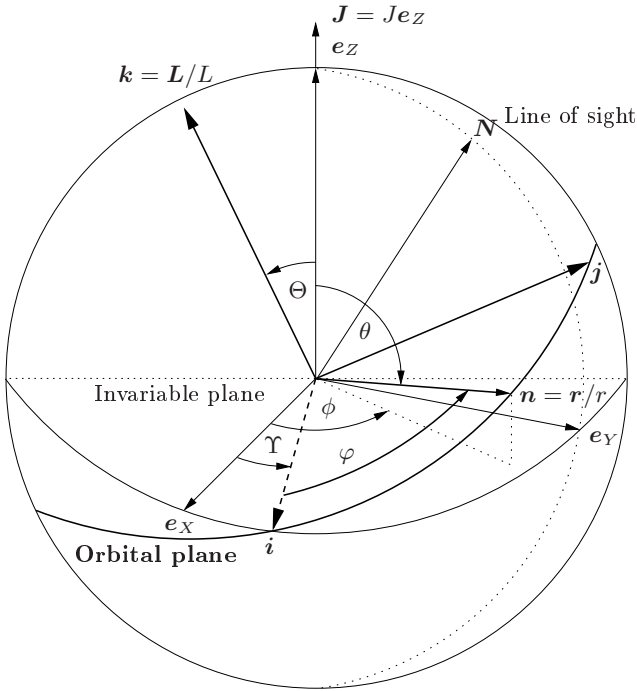}
  \caption{The binary geometry and interpretations of various angles
  appearing in this section.
  Our reference frame is ($\vek{e}_{X}, \vek{e}_{Y},\vek{e}_{Z}$),
  where the basic vector $\vek{e}_{Z}$ is aligned with the fixed total
  angular momentum vector $\vek{J} = \vek{L} + \vek{S}$,
  such that $\vek{J} = J \vek{e}_Z$.
  The invariable plane ($\vek{e}_{X}$, $\vek{e}_{Y}$) is
  perpendicular to $\vek{J}$. Important for the observation is the
  line--of--sight unit vector $\vek{N}$ from the observer to the
  source (compact binary).
  We may have, by a clever choice of $\vek{e}_{X}$ and $\vek{e}_{Y}$, the
  line--of--sight unit vector $\vek{N}$ in the
  $\vek{e}_{Y}$--$\vek{e}_{Z}$--plane. 
  $\vek{k} = \vek{L}/L$ is the unit vector in the direction of the
  orbital angular momentum $\vek{L}$, which is perpendicular to the
  orbital plane. The constant inclination of the orbital plane with
  respect to the invariable plane is $\Theta$, which is also the
  precession cone angle of $\vek{L}$ around $\vek{J}$.
  The orbital plane intersects the invariable plane at the line of
  nodes $\vek{i}$, with the longitude $\Upsilon$ measured in the
  invariable plane from $\vek{e}_{X}$.
  $\Upsilon$ is also the phase of the orbital plane precession.
  The orbital plane is spanned by the basic vectors
  ($\vek{i}, \vek{j}$), where $\vek{j} = \vek{k} \times \vek{i}$.
  }
  \label{fig:sphere_xyz_ijk} 
\end{figure}

Let us now find parametric solution to the angular parts of the
orbital equations of motion, Eqs.~\eqref{eq:eom_newton_with_so} combined with Eq.~\eqref{eq:Seff_eqauls_coupling_constant_S},
written in the form
\begin{subequations}
\label{eq:eom_second_and_third_again}
\begin{align}
\label{eq:eom_second_again}
r \dot{\theta}
&= p_\theta \left( 1 - \frac{\chi_\text{so}}{c^2 r} \right)
\,,
\\
\label{eq:eom_third_again}
r \sin\theta \dot{\phi} 
&= p_\phi \left( 1 - \frac{\chi_\text{so}}{c^2 r} \right) 
+ \frac{\chi_\text{so} J \sin\theta}{c^2 r^2}
\,.
\end{align}
\end{subequations}
For the cases of interest, where $L_{Z}$ is a constant, we may
introduce a constant angle $\Theta$ between $\vek{e}_{Z}$ and
$\vek{L}$ such that $L_{Z} = L \cos \Theta$
(see Fig.~\ref{fig:sphere_xyz_ijk}) and Eqs.~\eqref{eq:comp_pphi_of_p}
and \eqref{eq:comp_ptheta_of_p} simplify to
\begin{subequations}
\label{eq:p_phi_with_Theta_AND_p_theta2_with_Theta}
\begin{align}
\label{eq:p_phi_with_Theta}
p_\phi 
&= \frac{L}{r} \frac{\cos\Theta}{\sin\theta}
\,,
\\
\label{eq:p_theta2_with_Theta}
p_{\theta}^2 
&= \frac{L^2}{r^2} \left(1 - \frac{\cos^2 \Theta}{\sin^2 \theta} \right)
\,.
\end{align}
\end{subequations}

The efficient way of determining solution to the
Eqs.~\eqref{eq:eom_second_and_third_again}
begins by expressing the radial vector $\vek{r}$ in terms of an orbital triad.
This is achieved by connecting via two rotations the reference triad 
$(\vek{e}_{X}, \vek{e}_{Y},\vek{e}_{Z})$ to an orbital triad defined by
$(\vek{i}, \vek{j}, \vek{k})$.
In our orbital orthonormal triad, the unit vector $\vek{k}$, given by
$\vek{k} = \vek{L}/L$, and it along with $\vek{e}_{Z}$ defines
\begin{align}
\label{eq:def_line_of_nodes_i}
\vek{i} &= \frac{\vek{e}_Z \times \vek{k}}{|\vek{e}_Z \times \vek{k}|}
\end{align}
Note that $|\vek{e}_Z \times \vek{k}| = \sin \Theta$, which is
indeed always positive because of $0 \le \Theta < \pi$.
We note that the unit vector $\vek{i}$ gives the line of nodes,
associated with the intersection of the orbital plane with 
the invariable plane $(\vek{e}_{X},\vek{e}_{Y})$. The corresponding
inclination angle is $\Theta$.
Since the above mentioned line of nodes vanishes in the non--spinning
limit, the unit vector $\vek{i}$ is only defined when spin effects 
are included in the binary dynamics.

The two rotations arise from the observation that the (instantaneous)
position and orientation of the orbital plane 
with respect to the reference triad
$(\vek{e}_{X}, \vek{e}_{Y}, \vek{e}_{Z})$ are defined by two angles:
the longitude of the line of nodes 
$\Upsilon$, $(0 \le \Upsilon < 2 \pi)$ 
and the inclination
of the orbital plane with respect to the invariable
$\vek{e}_{X}$--$\vek{e}_{Y}$--plane $\Theta$, $(0 \le \Theta < \pi)$.
In terms of rotational matrices, we have
\begin{align}
\begin{pmatrix}
\vek{i}\\
\vek{j}\\
\vek{k}
\end{pmatrix}
&=
\begin{pmatrix}
1&0&0\\
0&\cos \Theta&\sin \Theta\\
0&-\sin \Theta&\cos \Theta
\end{pmatrix}
\begin{pmatrix}
\cos\Upsilon&\sin\Upsilon&0\\
-\sin\Upsilon&\cos\Upsilon&0\\
0&0&1
\end{pmatrix}
\begin{pmatrix}
\vek{e}_{X}\\
\vek{e}_{Y}\\
\vek{e}_{Z}
\end{pmatrix}
\,.
\nonumber
\end{align}

In the new orbital orthonormal triad
$(\vek{i}, \vek{j}, \vek{k})$,
the relative separation vector $\vek{r}$ is given by
\begin{align}
\label{eq:vector_r_in_ijk_frame}
\vek{r} &=
r \cos\varphi \, \vek{i}
+ r \sin\varphi \, \vek{j}
\,,
\end{align}
where the unit vectors 
$(\vek{i}, \vek{j}, \vek{k})$ in terms of
the reference triad $(\vek{e}_{X}, \vek{e}_{Y}, \vek{e}_{Z})$
are explicitly given by
\begin{subequations}
\label{eq:ijk_explicitly}
\begin{align}
\vek{i} &=
\cos\Upsilon \, \vek{e}_{X}
+ \sin\Upsilon \, \vek{e}_{Y}
\,,
\\
\vek{j} &=
- \cos\Theta \sin\Upsilon \, \vek{e}_{X}
+ \cos\Theta \cos\Upsilon \, \vek{e}_{Y}
+ \sin\Theta \, \vek{e}_{Z}
\,,
\\
\label{eq:k_parametrization_with_upsilon}
\vek{k} &=
\sin\Theta \sin\Upsilon \, \vek{e}_{X}
-\sin\Theta \cos\Upsilon \, \vek{e}_{Y}
+\cos\Theta \, \vek{e}_{Z}
\,,
\end{align}
\end{subequations}
The geometrical interpretations of the newly introduced angles and
basic vectors are illustrated in Fig.~\ref{fig:sphere_xyz_ijk}.

The comparison of $\vek{r}$, given by Eqs.~\eqref{eq:vec_r_in_r_theta_phi} and 
\eqref{eq:vec_n_in_r_theta_phi}, with
the one in the new angular variables, Eq.~\eqref{eq:vector_r_in_ijk_frame} with Eqs.~\eqref{eq:ijk_explicitly}, implies the following transformations 
for the angular variables: 
\begin{equation}
\label{eq:transfo_angular_coordinates}
(\theta, \phi) \longrightarrow (\Upsilon, \varphi):
\begin{cases}
\cos \theta = \sin \varphi \sin \Theta
\\
\sin(\phi-\Upsilon) \sin \theta = \sin \varphi \cos \Theta
\\
\cos(\phi-\Upsilon) \sin \theta = \cos \varphi
\end{cases}
\end{equation}
A clever combination of the above angular transformation equations and
their corresponding time derivatives leads to the following relations
for the angular velocities: 
\begin{subequations}
\label{eq:transf_angular_velocities}
\begin{align}
\label{eq:phi_dot_square}
\dot{\theta}^2 &
= \left(1 - \frac{\cos^2 \Theta}{\sin^2 \theta} \right) {\dot{\varphi}}^2
\,,
\\
\label{eq:phi_dot}
\dot{\phi} &= \dot{\Upsilon} + \frac{\cos\Theta}{\sin^2 \theta} \dot{\varphi} 
\,,
\end{align}
\end{subequations}
where the over dot again means the derivative with respect to time $t$
and $\Theta$ is treated as a constant angle.

Using Eq.~\eqref{eq:phi_dot_square} for ${\dot{\theta}}^2$ in Eq.~\eqref{eq:eom_second_again} with \eqref{eq:p_theta2_with_Theta} leads to
\begin{align}
\label{eq:dot_varphi_in_L_r_c_Xso}
\dot{\varphi} = \frac{L}{r^2} \left(1 - \frac{\chi_\text{so}}{c^2 r} \right)
\,.
\end{align}
Note that the above derivation requires $\Theta \neq 0$, which is
always satisfied for $\vek{S} \neq 0$, and hence
Eq.~\eqref{eq:dot_varphi_in_L_r_c_Xso} is not defined when $\Theta = 0$.
We again invoke the procedure employed in Sec.~III of
Ref.~\cite{MGS} to obtain the following 
parametric solution for $\varphi$:
\begin{subequations}
\label{eq:varphi_varphi0}
\begin{align}
\label{eq:varphi_varphi0_1}
\varphi - \varphi_{0} &= ( 1 + k ) v
\,,
\\
\label{eq:varphi_varphi0_2}
v &= 2 \arctan \left[ \left( \frac{1+e_\varphi}{1-e_\varphi}
\right)^{1/2} \tan \frac{u}{2} \right]
\,,
\end{align}
\end{subequations}
where $\varphi_0$ is the value of $\varphi$ at the time $t_0$.
The quantity $k$ is a measure of the advance of the periastron and 
$e_\varphi$ is the so called angular eccentricity.
They are expressible in terms of $E, L$ and 
$\vek{L} \cdot \vek{S}_\text{eff}$ as
\begin{subequations}
\begin{align}
k &= \frac{1}{c^2 L^2} 
\left(\chi_\text{so} - 3 \frac{\vek{L} \cdot \vek{S}_\text{eff}}{L^2} \right)
\,,
\\
e_\varphi^2 
&= 1 + 2 E L^2 - 4 \chi_\text{so} ( 1 + 2 E L^2) \frac{E}{c^2}
\nonumber
\\
& \quad
+ 4 ( 3 + 4 E L^2) \frac{\vek{L} \cdot \vek{S}_\text{eff}}{L^2} \frac{E}{c^2}
\,.
\end{align}
\end{subequations}
The parametrization for $r$, $l$ and $\varphi$ can also be obtained using the
\emph{conchoidal} transformations employed in Ref.~\cite{Timing_S_DD85}
and found to be in total agreement.

We now move on to derive the time evolution of $\Upsilon$, 
the longitude of the line of intersection 
(denoted by the line of nodes unit vector $\vek{i}$ in Fig.~\ref{fig:sphere_xyz_ijk}),
in a parametric way. 
Using Eq.~\eqref{eq:phi_dot} for $\dot{\phi}$ 
with Eq.~\eqref{eq:dot_varphi_in_L_r_c_Xso}
in Eq.~\eqref{eq:eom_third_again} with Eq.~\eqref{eq:p_phi_with_Theta}
we get the following differential equation for $\Upsilon$:
\begin{align}
\label{eq:diff_eq_dupsilon_dt}
\frac{d \Upsilon}{dt} &
= \frac{\chi_\text{so} J}{c^2 r^3}
\,,
\end{align}
which is also restricted to the spinning case $(\Theta \neq 0)$, 
since we made use of Eq.~\eqref{eq:dot_varphi_in_L_r_c_Xso} for $\dot{\varphi}$.

Note that the r. h. s. of the above equation is identical
to $\omega_p$, given by Eq.~\eqref{eq:definition_of_omega_p}.
This is not surprising as the frequencies of the precessional motion for 
$\vek{k} = \vek{L}/L$ and $\vek{i}$ should be identical due to
Eqs.~\eqref{eq:def_line_of_nodes_i} and \eqref{eq:ijk_explicitly}.
Further, a close inspection of Fig.~\ref{fig:sphere_xyz_ijk} 
reveals that the phase of the projection of $\vek{k}$ on to
the invariable plane $(\vek{e}_X,\vek{e}_Y)$,
as measured from $\vek{e}_X$, is given by $\Upsilon + 270^{\circ}$.

To solve the differential equation~\eqref{eq:diff_eq_dupsilon_dt}
analytically, we divide it 
by $d\varphi /dt$, given by Eq.~\eqref{eq:dot_varphi_in_L_r_c_Xso},
and deduce at the leading order
\begin{align}
\label{eq:diff_eq_dupsilon_dvarphi}
\frac{d \Upsilon}{d \varphi} = \frac{\chi_\text{so} J}{L} \frac{1}{c^2 r}
\,.
\end{align}
The radial separation $r$ appearing on the r. h. s. of above equation may be replaced by following expression in terms of $v$ as shown below
\begin{align}
\label{eq:r_newton_with_v}
r &= a (1 - e \cos u )
=\frac{a (1 - e^2)}{1 + e \cos v }
=\frac{L^2}{1 + e \cos v }
\,,
\end{align}
where $a$ and $e$ may be treated as the Newtonian accurate expressions
for $a_r$ and $e_r$.

This allows us to write Eq.~\eqref{eq:diff_eq_dupsilon_dvarphi} 
--- within the accuracy needed --- as
\begin{align}
\label{eq:dUpsilon_dv}
\frac{d \Upsilon}{dv} = \frac{\chi_\text{so} J}{c^2 L^3} ( 1 + e \cos v )
\,,
\end{align}
where we also used the fact that $\varphi = v$ at the leading order.
The Eq.~\eqref{eq:dUpsilon_dv} is easily integrated to 
obtain the parametric solution to $\Upsilon$ as 
\begin{align}
\label{eq:solution_upsilon_in_v}
\Upsilon - \Upsilon_{0} =  \frac{\chi_\text{so} J}{c^2 L^3} ( v + e \sin v )
\,,
\end{align}
where we put $v(t_{0}) = 0$.

There exists an alternate way to obtain
$d\Upsilon/dt$, given by Eq.~\eqref{eq:diff_eq_dupsilon_dt}.
Noting that the time evolution for the unit vector $\vek{k}$,
defined by Eq.~\eqref{eq:k_parametrization_with_upsilon},
is solely given by that for $\Upsilon$,
we compute the time derivative of $\vek{k}$ and it reads
\begin{align}
\label{eq:dot_k_with_upsilon}
\dot{\vek{k}} &=
\dot{\Upsilon}\sin\Theta (\cos\Upsilon \vek{e}_{X}
+\sin\Upsilon \vek{e}_{Y})
\,.
\end{align}
Using the fact that $\vek{L} = L \vek{k}$ and 
$\dot{\vek{L}} = L \dot{\vek{k}}$ in Eq.~\eqref{eq:L_precess_about_J},
we arrive (again) at $d\Upsilon/dt$, as given by
Eq.~\eqref{eq:diff_eq_dupsilon_dt}. 
Therefore, the parametric solution for $\Upsilon$
readily defines a similar solution for $\vek{L} = L \vek{k}$,
where $\vek{k}$ is given by 
Eqs.~\eqref{eq:k_parametrization_with_upsilon} with
\eqref{eq:solution_upsilon_in_v}.
The parametric solution to $\vek{L}$ reads
\begin{align}
\vek{L} &
= L (\sin\Theta \sin\Upsilon \, \vek{e}_{X}
- \sin\Theta \cos\Upsilon \, \vek{e}_{Y}
+ \cos\Theta \, \vek{e}_{Z})
\,.
\end{align}

Finally, let us derive a parametric solution to the precessional 
motion of $\vek{S}$, given by Eq.~\eqref{eq:S_precess_about_J}.
This parametric solution follows immediately by noting that 
$\vek{S} = \vek{J} - \vek{L}$, 
where $\vek{J} = J \vek{e}_Z$ and $\vek{L} = L \vek{k}$.
Using Eq.~\eqref{eq:k_parametrization_with_upsilon} for $\vek{k}$, we 
deduce that 
\begin{align}
\label{eq:spin_S_with_upsilon}
\vek{S}
&= L \sin\Theta (-\sin\Upsilon) \vek{e}_{X}
+ L \sin\Theta \cos\Upsilon \vek{e}_{Y}
\nonumber
\\
&\quad
+ (J - L \cos\Theta) \vek{e}_{Z}
\,,
\end{align}
The above equation, along with the parametric solution for 
$\Upsilon$, describes, in a parametric way, the time evolution of $\vek{S}$.

We finally collect all the relevant equations,
namely Eqs.~\eqref{r_and_n(t_t0)},
\eqref{eq:varphi_varphi0}, and \eqref{eq:solution_upsilon_in_v},
and display below our \emph{parametric solution for the binary
dynamics} given by ${H}_\text{NSO}$:
\begin{align}
\label{eq:solution_for_r_only_so}
\vek{r} &= r \cos\varphi \, \vek{i} + r \sin\varphi \, \vek{j}
\,,
\\
\vek{L} &= L \vek{k}
\,,
\\
\vek{S} &= J \vek{e}_{Z} - L \vek{k}
\,,
\end{align}
where the basic vectors $(\vek{i}, \vek{j}, \vek{k})$ are
explicitly given by 
\begin{subequations}
\begin{align}
\vek{i} &=
\cos\Upsilon  \vek{e}_{X}
+ \sin\Upsilon  \vek{e}_{Y}
\,,
\\
\vek{j} &=
- \cos\Theta \sin\Upsilon  \vek{e}_{X}
+ \cos\Theta \cos\Upsilon  \vek{e}_{Y}
+ \sin\Theta \, \vek{e}_{Z}
\,,
\\
\vek{k} &=
\sin\Theta \sin\Upsilon  \vek{e}_{X}
-\sin\Theta \cos\Upsilon  \vek{e}_{Y}
+\cos\Theta \, \vek{e}_{Z}
\,,
\end{align}
\end{subequations}
The time evolution for the radial and angular variables is given by
\begin{subequations}
\begin{align}
r &= a_r ( 1 - e_r \cos u )
\,,
\\
n(t-t_0) &= u - e_t \sin u
\,,
\\
\varphi-\varphi_{0} &= ( 1 + k ) v
\,,
\\
\label{eq:Upsilon_solution_in_sec4}
\Upsilon - \Upsilon_0 &= \frac{\chi_\text{so} J}{c^2 L^3} ( v + e \sin v )
\,,
\\
v &= 2 \arctan \left[ \left( \frac{1 + e_\varphi}{1 - e_\varphi}
\right)^{1/2} \tan \frac{u}{2} \right]
\,,
\end{align}
\end{subequations}
where the orbital elements are given by 
\begin{subequations}
\label{eq:parameters_for_NSO}
\begin{align}
a_r &= -\frac{1}{2 E}\left(1 - 2 \chi_\text{so} \cos\alpha \frac{S}{L}
\frac{E}{c^2}\right) 
\,,
\\
e_r^2 &= 1 + 2 E L^2 + 8 (1 + E L^2) \chi_\text{so} \cos\alpha \frac{S}{L} \frac{E}{c^2}
\,,
\\
n &= (-2 E)^{3/2}
\,,
\\
e_t^2 &= 1 + 2 E L^2 + 4 \chi_\text{so} \cos\alpha \frac{S}{L}\frac{E}{c^2}
\,,
\\
k & = \frac{1}{c^2 L^2} \left( \chi_\text{so} 
- 3 \chi_\text{so}\cos\alpha \frac{S}{L} \right)
\
\,,
\\
e_\varphi^2 &= 1 + 2 E L^2 - 4 ( 1 + 2 E L^2) \chi_\text{so} \frac{E}{c^2}
\nonumber \\
& \quad
+ 4 ( 3 + 4 E L^2) \chi_\text{so} \cos\alpha \frac{S}{L} \frac{E}{c^2}
\,.
\end{align}
\end{subequations}
As noted earlier, $e$ is given by the Newtonian contribution to $e_r$.
In Eqs.~\eqref{eq:parameters_for_NSO}, we have replaced
$\vek{L} \cdot \vek{S}_\text{eff}$ by $\chi_\text{so} L S \cos\alpha$, where
$\alpha$ is the constant angle between $\vek{L}$ and $\vek{S}$ and
$\chi_\text{so}$ is given by
\begin{equation}
\label{eq:coupling_constant_in_cases_again}
\chi_\text{so} :=
\begin{cases}
\delta_{1} = \delta_{2} = 7/8
& \text{for (i), the equal--mass--case}
\,,
\\
\delta_{1} \; \text{or} \; \delta_{2}
& \text{for (ii), the single--spin--case}
\,.
\end{cases}
\nonumber
\end{equation}
Further, we note that the constant angle $\Theta$ is not a free
variable and is given by one of the following relations:
\begin{subequations}
\label{eq:theta_in_terms_of_SLJ}
\begin{align}
\sin\Theta &= \frac{S \sin\alpha}{J}
\,,
\\
\cos\Theta &= \frac{L + S \cos\alpha}{J}
\,,
\end{align}
\end{subequations}
where the magnitude of the (reduced) total angular momentum is given by
$J = (L^2 + S^2 + 2 L S \cos\alpha)^{1/2}$.

We are aware that the above parametrization does not lead 
to the classic Keplerian parametrization simply by putting $S = 0$.
There are several arguments for this apparent lack of
a simple non--spinning limit.
Notice that $\Upsilon$ and $\varphi$ are defined 
with respect to the line of intersection,
characterized by the line of nodes unit vector $\vek{i}$. 
When the spin effects are neglected, the above line of intersection 
disappears, since the orbital plane becomes the invariable plane 
and the associated inclination angle $\Theta$ vanishes.
In this case, the related angles $\Upsilon$ and $\varphi$ 
are not individually defined,
though Eq.~\eqref{eq:transfo_angular_coordinates} 
gives us $\varphi + \Upsilon = \phi$, as required.
We emphasize that the related differential equations,
given by Eqs.~\eqref{eq:dot_varphi_in_L_r_c_Xso} and
\eqref{eq:diff_eq_dupsilon_dt}, are obtained using
$\Theta \neq 0$, corresponding to $S \neq 0$. Therefore,
Eqs.~\eqref{eq:dot_varphi_in_L_r_c_Xso} and
\eqref{eq:diff_eq_dupsilon_dt} are not defined if $\Theta = 0$.
Further, the apparent non--vanishing of the time evolution
of $\Upsilon$ in the non--spinning limit is also attributable
to the similar behavior for $\omega_p$.
However, a close scrutiny of 
Eqs.~\eqref{eq:r_dot_square_in_E_L_LdotSeff},
\eqref{eq:eom_second_and_third_again} with
\eqref{eq:p_phi_with_Theta_AND_p_theta2_with_Theta},
\eqref{eq:transfo_angular_coordinates} and
\eqref{eq:transf_angular_velocities} for 
$S \rightarrow 0$, which implies $\Theta \rightarrow 0$ 
and $\theta \rightarrow \pi /2$, 
reveals that the binary dynamics in the non--spinning case
is describable in terms of $r$ and $\phi$, where $\phi = \varphi + \Upsilon$,
as explained earlier.

Our parametric solution generalizes a restricted 
analysis considered in Ref.~\cite{Wex1995}. The
analysis, given in Ref.~\cite{Wex1995}, neglects the precessional motion of
the spin vectors and restricts $\vek{S}_\text{eff}$, denoted by
$\vek{\zeta}$ in Ref.~\cite{Wex1995}, to lie along $\vek{e}_{Z}$.
Due to these restrictions, that analysis 
provides only a parametric solution to
$\vek{r}$. We have, in the other hand, consistently taken into
consideration all the leading order spin--orbit interactions and
obtained a parametric solution to the entire binary dynamics.
In a future communication we will connect the above presented
parametric solution to quantities related to observation,
especially associated with binary pulsars 
\cite{KG2004timing}. 

In the next section, we will combine the Keplerian--type
parametrization developed above with the one presented in Sec.~\ref{Sec3}.

\section{Third post--Newtonian accurate generalized quasi--Keplerian
parametrization for compact binaries in eccentric orbits
with the first--order spin--orbit interaction}
\label{Sec5}

We are now in a position to have \emph{parametric solution to the dynamics},
defined by the Hamiltonian ${H}$, as given by Eq.~\eqref{H_3}.
This is done by combining consistently the parametrizations presented
in the previous two sections, Sec.~\ref{Sec3} and Sec.~\ref{Sec4}.
The two parametric solutions of Sec.~\ref{Sec3} and Sec.~\ref{Sec4} can be added
linearly to obtain the PN accurate binary dynamics, given by
Eqs.~\eqref{H_3} and \eqref{H_3_Full}, for the following reasons.
Recall that in the Hamiltonian, given by Eqs.~\eqref{H_3} and \eqref{H_3_Full},
the spin--orbit contribution is added linearly to the PN 
accurate non--spinning contributions. Moreover, as we are considering 
only leading order spin--orbit interactions, spin--orbit contributions
only cross with the Newtonian terms in the Hamiltonian.
This is why we can treat the spin--orbit contributions and the 
non--spinning PN contributions separately and later 
add the two parametric solutions linearly.
However, as a cautionary note, we state that care should be taken
while merging the above two parametric solutions to avoid adding
similar contributions twice.
Finally, we display below, in its
entirety, the parametric solution to the conservative third
post--Newtonian dynamics of 
spinning compact binaries moving in an eccentric orbit, in ADM--type
coordinates, with spin effects 
restricted to the leading order spin--orbit interactions:
\begin{align}
\label{eq:solution_for_r_3PN_plus_SO}
\vek{r} &= r \cos\varphi \, \vek{i} + r \sin\varphi \, \vek{j}
\,,
\\
\vek{L} &= L \vek{k}
\,,
\\
\vek{S} &= J \vek{e}_{Z} - L \vek{k}
\,,
\end{align}
where the basic vectors $(\vek{i}, \vek{j}, \vek{k})$ are
explicitly given by 
\begin{subequations}
\label{eq:ijk_in_the_combined_sol}
\begin{align}
\vek{i} &=
\cos\Upsilon  \vek{e}_{X}
+ \sin\Upsilon  \vek{e}_{Y}
\,,
\\
\vek{j} &=
- \cos\Theta \sin\Upsilon  \vek{e}_{X}
+ \cos\Theta \cos\Upsilon  \vek{e}_{Y}
+ \sin\Theta \, \vek{e}_{Z}
\,,
\\
\vek{k} &=
\sin\Theta \sin\Upsilon  \vek{e}_{X}
-\sin\Theta \cos\Upsilon  \vek{e}_{Y}
+\cos\Theta \, \vek{e}_{Z}
\,,
\end{align}
\end{subequations}
and
\begin{widetext}
\begin{subequations}
\label{e:FinalParamADM_3PN_plus_SO}
\begin{align}
r &= a_r \left( 1 -e_r \cos u \right )
\,,
\\
\label{eq:kepler_eq_in_combined_solution}
l \equiv n \left( t - t_0 \right)
&= u - e_t \sin u + \left( \frac{g_{4t}}{c^4}
+ \frac{g_{6t}}{c^6} \right) (v - u)
+ \left( \frac{f_{4t}}{c^4} + \frac{f_{6t}}{c^6} \right) \sin v
+ \frac{i_{6t}}{c^6} \sin 2 v
+ \frac{h_{6t}}{c^6}  \sin 3 v
\,,
\\
\varphi - \varphi_{0}
&= (1 + k) v
+ \left( \frac{f_{4\varphi}}{c^4}
+ \frac{f_{6\varphi}}{c^6} \right) \sin 2 v
+ \left( \frac{g_{4\varphi}}{c^4}
+ \frac{g_{6\varphi}}{c^6} \right) \sin 3 v
+ \frac{i_{6\varphi}}{c^6} \sin 4 v
+ \frac{h_{6\varphi}}{c^6} \sin 5 v
\,,
\\
\label{eq:Ups_min_Ups0_in_the_comb_sol}
\Upsilon - \Upsilon_0 &= \frac{\chi_\text{so} J}{c^2 L^3} ( v + e \sin v )
\,,
\\
\text{where} \quad
v &= 2\arctan \left[ \left( \frac{ 1 + e_{\varphi}}{ 1 - e_{\varphi}}
\right)^{1/2} \tan \frac{u}{2} \right]
\,.
\end{align}
\end{subequations}
The post--Newtonian accurate expressions for the orbital elements
$a_r$, $e_r^2$, $n$,
$e_t^2$, $k$, and $e_{\varphi}^2$
and the post--Newtonian orbital functions
$g_{4t}$, $g_{6t}$, $f_{4t}$, $f_{6t}$, $i_{6t}$, $h_{6t}$, $f_{4\varphi}$,
$f_{6\varphi}$, $g_{4\varphi}$, $g_{6\varphi}$, $i_{6\varphi}$,
and $h_{6\varphi}$, in terms of $E$, $L$, $S$, $\eta$ and $\alpha$, read
\begin{subequations}
\label{e:CoeffKP_3PN_with_SO}
\begin{align}
a_r &=
\frac{1}{ (-2 E) }
\bigg\{ 1 + \frac{ (-2 E) }{4 c^2}
\left( - 7 + \eta + 4 \chi_\text{so} \cos\alpha \frac{S}{L} \right)
+\frac{ (- 2 E)^2 }{ 16 c^4 }
\bigg[ 1 + 10 \eta + \eta^2
+ \frac{1}{ (-2 E L^2) } ( - 68 + 44 \eta)
\bigg] 
\nonumber
\\
& \quad
+ \frac{ (- 2 E)^3 }{ 192 c^6 }  
\biggl [ 3 - 9 \eta - 6 {\eta}^{2}
+3 {\eta}^{3} + \frac{1}{(-2 E L^2)}
\left ( 864 -3 {\pi}^{2} \eta - 2212 \eta + 432 {\eta}^{2} \right)
\nonumber
\\
& \quad
+\frac{1}{ (-2 E L^2)^2 }
\left(
- 6432 + 13488 \eta - 240 {\pi}^{2} \eta
- 768 {\eta}^{2} \right)
\biggr]   \bigg\}
\,,
\\
e_r^2
&= 1 + 2 E L^2 + \frac{(-2 E)}{4 c^2}
\biggl \{ 24 - 4 \eta - 5 (3-\eta) (-2 E L^2)
- 16 (1 + E L^2) \chi_\text{so} \cos\alpha \frac{S}{L} 
\biggr \} 
\nonumber
\\
& \quad  
+ \frac{ (-2 E)^2}{8 c^4}
\biggl \{
52 + 2 \eta + 2 {\eta}^{2}
- \left( 80-55 \eta+4 {\eta}^{2} \right) {(-2 E L^2)}
+\frac{8}{ (-2 E L^2) } ( 17 - 11 \eta)
\biggr \}
\nonumber
\\
& \quad 
+ \frac{ (-2 E)^3 }{ 192 c^6 }
\biggl\{
- 768 - 6 \pi^2 \eta 
- 344 \eta - 216 \eta^2 
+ 3 ( - 2 E L^2)
\left( - 1488 + 1556 \eta -319 \eta^2 + 4 \eta^3 \right)
\nonumber
\\
& \quad
- \frac{ 4 }{ (-2 E L^2)}
\left(588 - 8212 \eta + 177 \pi^2 \eta + 480 \eta^2 \right) 
+\frac{192}{ (-2 E L^2)^2 }
\left( 134 - 281 \eta + 5 \pi^2 \eta + 16 \eta^2 \right)  
\biggr\}
\,, 
\\
n &= (- 2 E)^{3/2} \bigg\{ 1 + \frac{ (-2 E) }{ 8 c^2 }  
\left( - 15 + \eta \right)
+ \frac{ (-2 E)^2 }{128 c^4 } 
\biggl[ 555 + 30 \eta + 11 \eta^2 
- \frac{ 192 }{ \sqrt{- 2 E L^2} } ( 5 - 2 \eta )
\biggr ]
\nonumber
\\
& \quad
+ \frac{ (-2 E)^3 }{ 3072 c^6 }
\biggl [- 29385 
-4995 \eta - 315 \eta^2 + 135 \eta^3
-\frac{16}{ (-2 E L^2)^{3/2} }
\left( 10080 + 123 \pi^2 \eta - 13952 \eta + 1440 \eta^2 \right)   
\nonumber
\\
& \quad
+ \frac{5760}{ \sqrt{-2 E L^2} }
\left(17 - 9 \eta + 2 \eta^2 \right)
\biggr] 
\bigg\}
\,, 
\\    
e_t^2
&= 1+ 2 E L^2 + \frac{ -2 E }{ 4 c^2 }
\bigg\{ 
- 8 + 8 \eta + (17 - 7 \eta ) (-2 E L^2) - 8 \chi_\text{so} \cos\alpha \frac{S}{L}
\bigg\} 
\nonumber
\\
& \quad
+ \frac{ (-2 E)^2 }{ 8 c^4 } \bigg\{ 8 + 4 \eta + 20 \eta^2
- (- 2 E L^2) ( 112 - 47 \eta + 16 \eta^2 )
+ 24 \sqrt{- 2 E L^2} (5 - 2 \eta) 
\nonumber
\\
& \quad    
+ \frac{4}{ (-2 E L^2) } \left( 17 - 11 \eta \right)
- \frac{24}{ \sqrt{ - 2 E L^2 } } \left( 5 - 2 \eta \right)
\bigg\}
\nonumber
\\
& \quad
+ \frac{ (-2 E)^3 }{ 192 c^6 }
\bigg\{ 24 \left( - 2 + 5 \eta \right) 
\left( - 23 + 10 \eta + 4 \eta^2 \right) - 15
\left( - 528 + 200 \eta - 77 \eta^2 + 24 \eta^3 \right) ( - 2 E L^2)
\nonumber
\\
& \quad
- 72 ( 265 - 193 \eta
+ 46 \eta^2 ) \sqrt{- 2 E L^2}
- \frac{2}{ ( - 2 E L^2) }
\left( 6732 + 117 \pi^2 \eta - 12508 \eta + 2004 \eta^2 \right)
\nonumber
\\
& \quad
+ \frac{2}{ \sqrt{- 2 E L^2} }
\left( 16380 - 19964 \eta + 123 \pi^2 \eta + 3240 \eta^2 \right)
- \frac{2}{ (-2 E L^2)^{3/2} }
\left( 10080 + 123 \pi^2 \eta - 13952 \eta + 1440 \eta^2 \right)
\nonumber
\\
& \quad
+ \frac{96}{ (-2 E L^2)^2 }
\left( 134 - 281 \eta + 5 \pi^2 \eta + 16 \eta^2 \right)
\bigg\}
\,,
\\
g_{4 t}
&= \frac{ 3 (-2 E)^2 }{2}
\frac{5 - 2 \eta }{ \sqrt{- 2 E L^2} }
\,, 
\\
g_{6 t}
&= \frac{ (-2 E)^3 }{192}
\biggl\{
\frac{1}{ (- 2 E L^2)^{3/2} }
\left( 10080 + 123 \pi^2 \eta - 13952 \eta + 1440 \eta^2 \right)
+ \frac{1}{ \sqrt{ -2 E L^2 } }
\left( - 3420 + 1980 \eta - 648 \eta^2 \right)
\biggr\}
\,,
\\ 
f_{4 t}
&= - \frac{1}{8} \frac{ (-2 E)^2 }{ \sqrt{-2 E L^2} }
(4 + \eta) \eta \sqrt{ 1 + 2 E L^2 }
\,,  
\\
f_{6 t}
&= \frac{ (- 2 E)^3 }{192}
\frac{1}{ \sqrt{1 + 2 E L^2} }
\bigg\{
\frac{1}{ (-2 E L^2)^{3/2} }  
( 1728 - 4148 \eta + 3 \pi^2 \eta + 600 \eta^2 + 33 \eta^3 )
+ 3 \sqrt{-2 E L^2} \eta ( - 64 - 4 \eta + 23 \eta^2 )
\nonumber
\\
& \quad
+ \frac{1}{ \sqrt{- 2 E L^2} }
( - 1728 + 4232 \eta - 3 \pi^2 \eta - 627 \eta^2 - 105 \eta^3 )
\bigg\}
\, , 
\\
i_{6 t}
&= \frac{(-2 E)^3}{32} \frac{ (1 + 2 E L^2) }{ (- 2 E L^2)^{3/2} } \eta
( 23 + 12 \eta + 6 \eta^2 ) 
\,,  
\\
h_{6 t}
&= \frac{ 13 (-2 E)^3 }{192} \eta^3
\biggl( \frac{ 1 + 2 E L^2 }{- 2 E L^2} \biggr)^{3/2}
\,, 
\\
k
&= \frac{3}{c^2 L^2}
\biggl\{
1 + \frac{\chi_\text{so}}{3} - \chi_\text{so} \cos\alpha \frac{S}{L}
+ \frac{ (-2 E) }{4 c^2}
\left( - 5 + 2 \eta + \frac{ 35 - 10 \eta }{ -2 E L^2 } \right)
+ \frac{ (-2 E)^2 }{ 384 c^4 } 
\biggl[
120 - 120 \eta + 96 \eta^2 
\nonumber
\\
& \quad
+ \frac{1}{ ( - 2 E L^2) }
( - 10080 + 13952 \eta - 123 \pi^2 \eta - 1440 \eta^2 )
\nonumber
\\
& \quad
+ \frac{1}{ (- 2 E L^2)^2 } 
( 36960 - 40000 \eta +  615 \pi^2 \eta + 1680 \eta^2 )
\biggr]
\biggr\}
\,,
\\
f_{ 4 \varphi}
&= \frac{ (-2 E)^2 }{8} \frac{ (1 + 2 E L^2) }{ ( - 2 E L^2)^2 } \eta
( 1 - 3 \eta)
\,,
\\
f_{6 \varphi}
&= \frac{1}{256} \frac{ (-2 E)^3 }{ (- 2 E L^2) }
\biggl\{
- ( 44 + 160 \eta - 96 \eta^2 ) \eta
+\frac{1}{ (- 2 E L^2) }
( - 256 - 49 \pi^2 \eta + 1096 \eta + 624 \eta^2 - 80 \eta^3 )
\nonumber
\\
& \quad+ \frac{1}{ (- 2 E L^2)^2 }
(256 + 49 \pi^2 \eta - 980 \eta - 672 \eta^2 - 40 \eta^3 )
\biggr\}
\,,
\\
g_{4 \varphi}
&= - \frac{3 (-2 E)^2 }{ 32 } \frac{ \eta^2 }{ (- 2 E L^2)^2 }
( 1 + 2 E L^2)^{3/2}
\,,
\\
g_{6 \varphi}
&= \frac{ (-2 E)^3 }{768} \frac{ \sqrt{1 + 2 E L^2} }{ (- 2 E L^2) } \eta
\biggl\{ - 27 \eta + 78 \eta^2  - \frac{ 1 }{ (- 2 E L^2) } 
( 220 + 3 \pi^2 + 96 \eta + 150 \eta^2 )
\nonumber
\\
& \quad
+ \frac{1}{ (- 2 E L^2)^2 } (220 + 3 \pi^2 - 120 \eta + 45 \eta^2 ) 
\biggr\}
\,,
\\
i_{6 \varphi}
&= \frac{ (-2 E)^3 }{128} \frac{ (1 + 2 E L^2)^2 }{ (- 2 E L^2)^3 } \eta 
( 5 + 28 \eta + 10 \eta^2 )
\,,
\\
h_{6 \varphi}
&= \frac{ 5 (- 2 E)^3 }{256} \frac{ \eta^3 }{ (-2 E L^2)^3 }
(1 + 2 E L^2)^{5/2}
\,,
\\
e_{\varphi}^2
&= 1 + 2 E L^2 + \frac{ (-2 E) }{ 4 c^2 }
\bigg\{24 - ( 15 - \eta ) (- 2 E L^2)
+ 8 ( 1 + 2 E L^2 ) \chi_\text{so}
- 8 ( 3 + 4 E L^2 ) \chi_\text{so} \cos\alpha \frac{S}{L}
\bigg\}   
\nonumber
\\
& \quad
+ \frac{ (-2 E)^2 }{ 16 c^4 }
\bigg\{
- 32 + 176 \eta + 18 \eta^2 - (- 2 E L^2) ( 160 - 30 \eta
+ 3 \eta^2 ) + \frac{1}{ (- 2 E L^2) } 
\left( 408 - 232 \eta - 15 \eta^2 \right)
\bigg\}
\nonumber
\\
& \quad
+ \frac{ (-2 E)^3 }{ 384 c^6} 
\bigg\{
- 16032 + 2764 \eta + 3 \pi^2 \eta + 4536 \eta^2 + 234 \eta^3
- 36 \left( 248 - 80 \eta + 13 \eta^2 + \eta^3 \right) (- 2 E L^2)
\nonumber
\\
& \quad
- \frac{ 6 }{ (- 2 E L^2) }
\left( 2456 - 26860 \eta + 581 \pi^2 \eta + 2689 \eta^2 + 10 \eta^3 \right)
\nonumber
\\
& \quad
+ \frac{ 3 }{ (-2 E L^2)^2 }
\left(27776- 65436 \eta + 1325 \pi^2 \eta + 3440 \eta^2 - 70 \eta^3 \right)
\biggr\}
\,.
\end{align}
\end{subequations}
We recall that $e$, appearing in
Eq.~\eqref{eq:Ups_min_Ups0_in_the_comb_sol}, is given by the 
Newtonian contribution to $e_r$. 
The three eccentricities $e_r$, $e_t$ and $e_\varphi$ are related to
each other by post--Newtonian corrections 
\begin{subequations}
\begin{align}
e_t 
&= e_r 
\bigg\{ 1 
+ \frac{ (-2 E) }{ 2 c^2 } 
\left( - 8 + 3 \eta + 2 \chi_\text{so} \cos\alpha \frac{S}{L} \right)
+ \frac{ (-2 E)^2 }{ 8 c^4 } 
\frac{1}{ (- 2 E L^2) } 
\bigg[ - 34 + 22 \eta
- ( 60 - 24 \eta ) \sqrt{ - 2 E L^2 }
\nonumber
\\
& \quad
+ ( 72 - 33 \eta + 12 \eta^2 ) (- 2 E L^2)
\bigg]
+ \frac{ (- 2 E)^3 }{ 192 c^6 } \frac{1}{ (- 2 E L^2)^2 }
\bigg[ - 6432 + 13488 \eta - 240 \eta \pi^2 - 768 \eta^2
\nonumber
\\
& \quad
+( - 10080 + 13952 \eta - 123 \eta \pi^2 - 1440 \eta^2 )
\sqrt{-2 E L^2}
+( 2700 - 4420 \eta - 3 \eta \pi^2 + 1092 \eta^2 ) (- 2 E L^2)
\nonumber
\\
& \quad
+( 9180 - 6444 \eta + 1512 \eta^2 ) ( - 2 E L^2 )^{3/2}
+ ( - 3840 + 1284 \eta - 672 \eta^2 + 240 \eta^3 ) ( - 2 E L^2 )^2
\bigg] \bigg\}
\,,
\\
e_\varphi 
& = e_r 
\bigg\{ 1
+ \frac{ (- 2 E) }{ 2 c^2 } 
\left( 
\eta + 2 \chi_\text{so} - 2 \chi_\text{so} \cos\alpha \frac{S}{L} 
\right) 
+ \frac{ (- 2 E)^2 }{ 32 c^4 } \frac{1}{ (- 2 E L^2 ) } 
\bigg[ 136 - 56 \eta - 15 \eta^2 + \eta ( 20 + 11 \eta ) (- 2 E L^2 ) 
\bigg]
\nonumber
\\
& \quad 
+ \frac{ (- 2 E)^3 }{ 768 c^6 } \frac{1}{ (- 2 E L^2 )^2 }
\bigg[ 31872 - 88404 \eta + 2055 \eta \pi^2 + 4176 \eta^2 - 210 \eta^3 
\nonumber
\\
& \quad 
+ ( 2256 + 10228 \eta - 15 \eta \pi^2 - 2406 \eta^2 - 450 \eta^3 )
(- 2 E L^2 )
+ 6 \eta ( 136 + 34 \eta + 31 \eta^2 ) (- 2 E L^2 )^2
\bigg] \bigg\}
\,,
\end{align}
\end{subequations}
\end{widetext}
and hence it is possible to
describe the binary dynamics in terms of one of the eccentricities.

We emphasize that, while adapting the parametrization to describe the
binary dynamics, care should be taken to restrict orbital elements to
the required post--Newtonian order.

As noted earlier, $\Theta$, the precessional angle for $\vek{k}$, is
given by
\begin{subequations}
\label{eq:theta_in_terms_of_SLJ_again}
\begin{align}
\sin\Theta &= \frac{S \sin\alpha}{J}
\,,
\\
\cos\Theta &= \frac{L + S \cos\alpha}{J}
\,,
\end{align}
\end{subequations}
where the magnitude of the total angular momentum is given by
$J = (L^2 + S^2 + 2 L S \cos\alpha)^{1/2}$.

Finally, we want to emphasize that the parametric solution, given by
Eqs.~\eqref{eq:solution_for_r_3PN_plus_SO}--\eqref{eq:theta_in_terms_of_SLJ_again},
describes the entire conservative post--Newtonian accurate
dynamics of a spinning compact binary in an eccentric orbit, 
when the spin effects are restricted to the leading order 
spin--orbit interaction. This means the parametrization consistently
describes not only the precessional motion of the orbit
inside the orbital plane, but also the precessional motions 
of the orbital plane and the spins themselves.

It is possible, though tedious, to show that 
the combined parametric solution, given by 
Eqs.~\eqref{eq:solution_for_r_3PN_plus_SO}--\eqref{eq:theta_in_terms_of_SLJ_again},
is indeed a parametric solution
to the total orbital dynamics prescribed by Eqs.~\eqref{H_3} and \eqref{H_3_Full}.
To show that, we have computed, using the combined parametric solution,
PN accurate expressions for $\dot{r}$, $\dot{\theta}$ and $\dot{\phi}$ 
in terms of $E$, $L$, $\vek{L} \cdot \vek{S}_\text{eff}$
and $r$ with the help of Eqs.~\eqref{eq:transfo_angular_coordinates} and \eqref{eq:transf_angular_velocities}. These expressions were found 
to be in total agreement with PN accurate expressions for 
$\dot{r}$, $\dot{\theta}$ and $\dot{\phi}$ computed 
from the total (reduced) Hamiltonian, given by 
Eqs.~\eqref{H_3} and \eqref{H_3_Full}, using the Hamilton's 
equations of motion
[See Eqs.~\eqref{eq:vec_n_in_r_theta_phi}--\eqref{eq:eom_newton_with_so}
and the related discussions].

\section{The leading part of the quadrupolar gravitational--wave
polarizations for spinning compact binaries in eccentric orbits}
\label{Sec6}

\begin{figure}[ht]
  \centering
  \psfrag{X}{$\vek{e}_X , \vek{p}$}
  \psfrag{Y}{$\vek{e}_Y$}
  \psfrag{Z}{$\vek{e}_Z$}
  \psfrag{J}{$\vek{J}= J \vek{e}_Z$}
  \psfrag{N}{$\vek{N}$}
  \psfrag{q}{$\vek{q}$}
  \psfrag{n}{$\vek{n} = \vek{r} / r$}
  \psfrag{iTheta}{$\vek{i}$}
  \psfrag{j}{$\vek{j}$}
  \psfrag{k}{$\vek{k}=\vek{L}/L$}
  \psfrag{i}{$i$}
  \psfrag{i0}{$i_{0}$}
  \psfrag{phi2}{$\varphi$}
  \psfrag{theta2}{$\Theta$}
  \psfrag{Upsilon}{$\Upsilon$}
  \psfrag{lineofsight}{\small{Line of sight}}
  \psfrag{peri}{\small{Periastron}}
  \psfrag{invariable}{\small{Invariable plane}}
  \psfrag{orbit}{\bf{\small{Orbital plane}}}
  \psfrag{sky}{\small{Plane of the sky}}
  \includegraphics[scale=0.65]{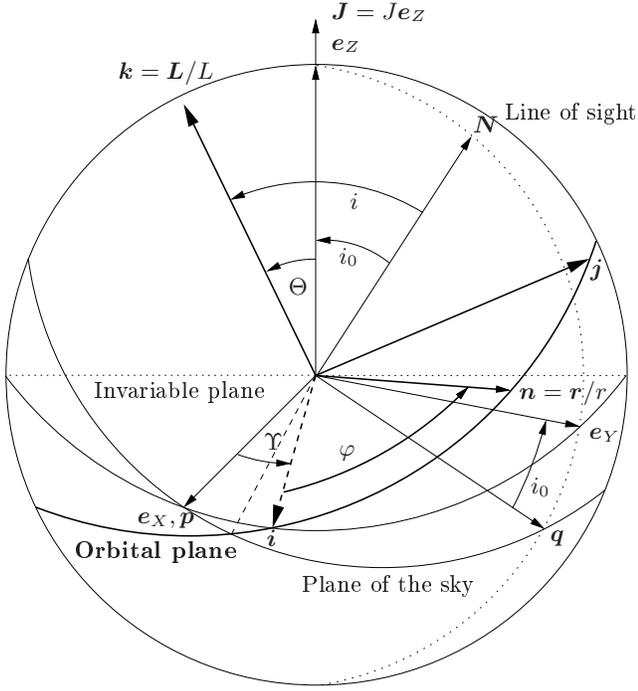} 
  \caption{
  The convention we adopted to link the 
  orbital frame $(\vek{i},\vek{j},\vek{k})$, the invariable frame
  $(\vek{e}_{X},\vek{e}_{Y},\vek{e}_{Z})$ and the 
  frame $(\vek{p},\vek{q},\vek{N})$ associated with the observer.
  Since we choose the line--of--sight unit vector $\vek{N}$ to lie in the
  $\vek{e}_{Y}$--$\vek{e}_{Z}$--plane, we may align the polarization
  vector $\vek{p}$ along $\vek{e}_{X}$, where the plane of 
  the sky meets the invariable plane. This implies that $\vek{q}$ 
  also lies in the $\vek{e}_{Y}$--$\vek{e}_{Z}$--plane.
  Therefore the frames $(\vek{p},\vek{q},\vek{N})$ and 
  $(\vek{e}_{X},\vek{e}_{Y},\vek{e}_{Z})$ are connected 
  by a constant angle $i_0$, which is the constant 
  inclination angle between $\vek{N}$ and $\vek{J}$.
  The polarization vectors $\vek{p}$ and $\vek{q}$ span the plane of
  the sky. The inclination of this plane with respect to the orbital
  plane is the \emph{orbital} inclination $i$.
  The inclination of the orbital plane with respect to the invariable
  plane is denoted by the constant angle $\Theta$.
  }
  \label{sphere_xyz_pqN} 
\end{figure}

As an application of the above parametrization, we obtain, for the
first time, explicit expressions for $h_{+}$ and $h_{\times}$, 
suitable to describe gravitational radiation from spinning compact
binaries moving in eccentric orbits, corresponding to the cases 
considered in Sec.~\ref{Sec4}.
The gravitational wave polarization states, $h_{+}$ and $h_{\times}$,
are usually given by 
\begin{subequations}
\label{eq:definition_hp_hx}
\begin{align}
h_{+} &= \frac{1}{2} \left(p_i p_j - q_i q_j \right) h_{ij}^\text{TT}
\,, \\
h_{\times} &= \frac{1}{2} \left(p_i q_j + p_j q_i \right) h_{ij}^\text{TT}
\,,
\end{align}
\end{subequations}
where $\vek{p}$ and $\vek{q}$ are two orthogonal unit vectors in the
plane of the sky, \emph{i.e.} a plane transverse to the radial
direction $\vek{N}$ (the line of sight), linking the observer to the
source (see Fig.~\ref{sphere_xyz_pqN}).

The transverse--traceless (TT) part of the radiation field,
$h_{ij}^\text{TT}$, which depends on the dynamics of the compact
binary, is expressible in terms of a post--Newtonian expansion
in $(v/c)$.
Symbolically, the TT radiation field may be written as
\begin{align}
\label{eq:symbolically_h_ij_expansion}
h_{ij}^\text{TT} &
= \frac{1}{c^4} h_{ij}^{(0)}
+ \frac{1}{c^5} h_{ij}^{(1)}
+ \frac{1}{c^6} h_{ij}^{(2)}
+ \frac{1}{c^7} h_{ij}^{(3)}
+ \cdots
\,.
\end{align}
In this paper, for simplicity, we will restrict
$h_{ij}^\text{TT}$ to its leading `quadrupolar' order, namely to
$h_{ij}^{(0)} /c^4$, and denote it by $h_{ij}^\text{TT}|_{\text Q}$.
However, higher PN corrections to $h_{ij}^\text{TT}$ are available in
the literature. The 2PN corrections to $h_{ij}^\text{TT}$ for compact
binaries moving in general orbits are given in Refs.~\cite{Will_Wiseman_1996,GI97}.
The spin--orbit and spin--spin interactions directly contribute to
$h_{ij}^\text{TT}$ and were computed in Ref.~\cite{K95}.
For compact binaries, in circular orbits, the explicit expressions for 2.5PN 
accurate $h_{ij}^\text{TT}$ --- corrections that include
$h_{ij}^{(5)} /c^9$ in harmonic coordinates --- were recently derived
in Ref.~\cite{ABIQ}. 
We recall that post--Newtonian corrections to $h_{ij}^\text{TT}$ are
usually given in harmonic coordinates, which differ from the one we
employed here at 2PN and 3PN orders. However, using coordinate
transformations that link harmonic and ADM--type coordinates, given in
Ref.~\cite{DD_S_part2}, it is possible to obtain
post--Newtonian corrections to
the amplitudes of $h_{+}$ and $h_{\times}$ in ADM--type coordinates.

The explicit expression for $h_{ij}^\text{TT}$, in the leading part of
the `quadrupolar' approximation, reads
\begin{align}
\label{eq:definition_h_newton}
h_{km}^\text{TT} \big|_{\text Q}
&= \frac{4 G \mu }{ c^4 R'} {\cal P}_{ijkm}(\vek{N})
\left( v_{ij} - \frac{G M}{r} n_{ij} \right)
\,,
\end{align}
where ${\cal P}_{ijkm}(\vek{N})$ is the usual transverse traceless
projection operator projecting vectors normal to $\vek{N}$, where
$\vek{N} = \vek{R'}/R'$ is the line--of--sight unit vector from the
observer to the binary, and $R' = |\vek{R'}|$ is the corresponding radial
distance. We also used $v_{ij} := v_i v_j$, and $n_{ij} := n_i n_j $,
where $v_i$ and $n_i$ are the components of the velocity vector
$\vek{v} = d \vek{r} / dt$ and the unit relative separation vector
$\vek{n} = \vek{r}/r$,  where $ r= |\vek{r}| $.

Using Eqs.~\eqref{eq:definition_hp_hx} and \eqref{eq:definition_h_newton}, 
we obtain $h_{+}|_{\text Q}$ and $h_{\times}|_{\text Q}$, the expressions for
gravitational wave polarizations when their amplitudes are restricted to the 
leading quadrupolar order, for compact binaries moving in general orbits as
\begin{subequations}
\label{eq:h_plus_and_h_cross_in_n_and_v}
\begin{align}
h_{+} \big|_{\text Q}
&= \frac{2 G \mu}{c^4 R'}
\left[ \left( p_i p_j - q_i q_j \right)
\left( v_{ij} - \frac{G M}{r} n_{ij} \right) \right]
\nonumber
\\
&= \frac{2 G \mu}{c^4 R'}
\bigg\{ 
(\vek{p} \cdot \vek{v} )^2 - (\vek{q} \cdot \vek{v} )^2
\nonumber
\\
&\quad
-  \frac{G M}{r} \left[ ( \vek{p} \cdot \vek{n} )^2
- ( \vek{q} \cdot \vek{n} )^2 \right] 
\bigg\}
\,, \\
h_{\times} \big|_{\text Q}
&= \frac{2 G \mu}{c^4 R'}
\left[ \left( p_i q_j + p_j q_i \right)
\left( v_{ij} - \frac{G M}{r} n_{ij} \right) \right]
\nonumber
\\
&= \frac{4 G \mu}{c^4 R'}
\left[ (\vek{p} \cdot \vek{v}) (\vek{q} \cdot \vek{v}) - \frac{G M}{r}
(\vek{p} \cdot \vek{n}) (\vek{q} \cdot \vek{n}) \right]
\,.
\end{align}
\end{subequations}

The expressions for $h_{+} |_{\text Q}$ and $h_{\times} |_{\text Q}$,
for spinning compact 
binaries in eccentric orbits are obtained by adapting the radial
separation and velocity vectors,
$\vek{r}$ and $\vek{v} =  d\vek{r}/dt$,
to the orthonormal triad $(\vek{p}, \vek{q}, \vek{N})$.
This is easily achieved in the following way.
We deduce that the explicit parametric representation for $\vek{r}$,
given by Eqs.~\eqref{eq:solution_for_r_3PN_plus_SO}
and \eqref{eq:ijk_in_the_combined_sol}, in terms of the reference triad
$(\vek{e}_{X}, \vek{e}_{Y}, \vek{e}_{Z})$ is given by
\begin{align}
\label{eq:vector_r_expanded}
\vek{r}
&=r (\cos\Upsilon \cos\varphi - \cos\Theta \sin\Upsilon \sin\varphi)
\vek{e}_{X}
\nonumber
\\
&\quad
+ r (\sin\Upsilon \cos\varphi + \cos\Theta \cos\Upsilon \sin\varphi)
\vek{e}_{Y}
\nonumber
\\
&\quad
+ r \sin\Theta \sin\varphi \,
\vek{e}_{Z}
\,.
\end{align}

In the above equation, the angles $\varphi$ and $\Upsilon$ along with $r$
depend on the orbital dynamics. However, $\Theta$ is a constant angle ---
a direct consequence of choosing two distinct cases for the 
parametrization (see discussion in Sec.~\ref{Sec4}).

To compute $h_{+} |_{\text Q}$ and $h_{\times} |_{\text Q}$,
one needs to choose a
convention for the direction and orientation of the orbit with respect
to the plane of sky.
In our approach, the orthonormal triad
$(\vek{p},\vek{q},\vek{N})$ is 
connected to the reference triad $(\vek{e}_{X}, \vek{e}_{Y},
\vek{e}_{Z})$ by a constant angle $i_0$
(see Fig.~\ref{sphere_xyz_pqN}).

In particular, we connect the reference triad $(\vek{e}_{X},
\vek{e}_{Y}, \vek{e}_{Z})$ to the observer triad $( \vek{p}, \vek{q},
\vek{N} )$ by a rotation around $ \vek{p}$ via a constant angle $i_{0}$
\begin{align}
\label{eq:transform_pqN_to_exeyez}
\begin{pmatrix}
\vek{e}_{X}\\
\vek{e}_{Y}\\
\vek{e}_{Z}
\end{pmatrix}
=
\begin{pmatrix}
1&0&0\\
0&C_{i_{0}} &S_{i_{0}} \\
0&-S_{i_{0}} &C_{i_{0}}
\end{pmatrix}
\begin{pmatrix}
\vek{p} \\
\vek{q} \\
\vek{N}
\end{pmatrix}
\,,
\end{align}
where $C_{i_{0}}$ and $S_{i_{0}}$ are shorthand notations for $\cos i_{0}$ and
$\sin i_{0}$. It is now straightforward to get the explicit expressions
for the radial separation and velocity vectors in terms of the orbital triad 
$(\vek{p}, \vek{q}, \vek{N})$, associated with the observer.
The vectorial equation for $\vek{r}$ reads
\begin{align}
\label{eq:vector_r_in_pqN}
\vek{r} &=
r (\cos\Upsilon \cos\varphi - C_\Theta \sin\Upsilon \sin\varphi)
\vek{p}
\nonumber\\
&\quad
+ r \left[C_{i_{0}} \sin\Upsilon \cos\varphi
- \left( S_{i_{0}} S_\Theta - C_{i_{0}} C_\Theta \cos\Upsilon \right)
\sin\varphi 
\right] \vek{q}
\nonumber\\
&\quad
+ r \left[S_{i_{0}} \sin\Upsilon \cos\varphi
+ \left( C_{i_{0}} S_\Theta + S_{i_{0}} C_\Theta \cos\Upsilon \right)
\sin\varphi 
\right] \vek{N}
\,,
\end{align}
where $C_{\Theta}$ and $S_{\Theta}$ are shorthand notations for
$\cos\Theta$ and $\sin\Theta$. 
The radial velocity vector $\vek{v} = d\vek{r}/dt$ is given by
\begin{widetext}
\begin{align}
\label{eq:vector_v_in_pqN}
\vek{v} &=
\left\{
\left[- C_\Theta \dot{r} \sin\Upsilon
- r (\dot{\varphi}+ C_\Theta \dot{\Upsilon}) \cos\Upsilon
\right]\sin\varphi 
+\left[- r (C_\Theta \dot{\varphi} + \dot{\Upsilon}) \sin\Upsilon
+ \dot{r} \cos\Upsilon
\right]\cos\varphi 
\right\}\vek{p}
\nonumber\\
&\quad
+\left\{
\left[-r C_{i_{0}} (\dot{\varphi} + C_\Theta \dot{\Upsilon}) \sin\Upsilon
+ C_{i_{0}} C_\Theta  \dot{r} \cos\Upsilon
- S_{i_{0}} S_\Theta  \dot{r} \right]\sin\varphi 
\right.
\nonumber\\
&\quad\quad
\left.
+\left[
C_{i_{0}} \dot{r} \sin\Upsilon
+r C_{i_{0}} (C_\Theta \dot{\varphi} + \dot{\Upsilon}) \cos\Upsilon
- S_{i_{0}} S_\Theta r \dot{\varphi}
\right]\cos\varphi 
\right\}\vek{q}
\nonumber\\
&\quad
+\left\{
\left[-r S_{i_{0}} (\dot{\varphi} + C_\Theta \dot{\Upsilon}) \sin\Upsilon
+ S_{i_{0}} C_\Theta \dot{r} \cos\Upsilon
+ C_{i_{0}} S_\Theta \dot{r} 
\right]\sin\varphi
\right.
\nonumber\\
&\quad\quad
\left.
+\left[S_{i_{0}} \dot{r} \sin\Upsilon
+ r S_{i_{0}} (C_\Theta \dot{\varphi} + \dot{\Upsilon}) \cos\Upsilon
+ C_{i_{0}} S_\Theta r \dot{\varphi}
\right]\cos\varphi 
\right\}
\vek{N}
\,.
\end{align}
Note that while computing $\vek{v}$, we have kept $\Theta$ and $i_{0}$
as constant angles. We point out that the way we defined the angle
$i_{0}$ is similar to the manner in which the `inclination'
was defined in Ref.~\cite{LBK1995} and
Ref.~\cite{K95} [see, e.g., discussions prior to Eqs.~(4.23) in
Subsec.~C under Sec.~IV in Ref.~\cite{K95}].
 
It is now straightforward, though lengthy, to obtain the explicit expressions 
for gravitational wave polarizations for spinning compact binaries
moving in eccentric orbits, when the dynamics includes
the leading spin--orbit interaction.
The expressions, $h_{+}|_{\text Q}$ and $h_{\times}|_{\text Q}$,
giving the gravitational wave polarizations with amplitudes due to 
the quadrupolar contributions, read
\begin{subequations}
\label{eq:h_plus_h_cross_with_SO}
\begin{align}
h_{+} \big|_{\text{Q}}
&=
\frac{2 G \mu}{c^4 R'}
\Bigg(
\bigg\{
-S_{i_{0}} C_{i_{0}} S_\Theta
\left( \frac{G M}{r} - \dot{r}^2 + r^2 \dot{\varphi}^2 + C_\Theta r^2
\dot{\varphi} \dot{\Upsilon}\right) \sin\Upsilon
+S_{i_{0}} C_{i_{0}} S_\Theta r \dot{r} (2 C_\Theta \dot{\varphi} +
\dot{\Upsilon}) \cos\Upsilon 
\nonumber \\
&\quad\quad
+\frac{1}{2} (1+C_{i_{0}}^2)
\left[ C_\Theta \left( \frac{G M}{r} - \dot{r}^2 + r^2 \dot{\varphi}^2 + r^2
\dot{\Upsilon}^2 \right) + (1+C_\Theta^2) r^2 \dot{\varphi}
\dot{\Upsilon} \right] \sin 2 \Upsilon
\nonumber \\
&\quad\quad
-\frac{1}{2} (1+C_{i_{0}}^2)
r \dot{r} \left[ (1+C_\Theta^2) \dot{\varphi} + 2 C_\Theta
\dot{\Upsilon} \right] \cos 2 \Upsilon
-\frac{3}{2} S_{i_{0}}^2 S_\Theta^2 r \dot{r} \dot{\varphi}
\bigg\} \sin 2 \varphi
\nonumber \\
&\quad
+\bigg\{
S_{i_{0}} C_{i_{0}} S_\Theta r \dot{r} (2 \dot{\varphi} + C_\Theta
\dot{\Upsilon}) 
\sin\Upsilon
+S_{i_{0}} C_{i_{0}} S_\Theta \left[  r^2 \dot{\varphi} \dot{\Upsilon}
+ C_\Theta \left( \frac{G M}{r} - \dot{r}^2 + r^2 
\dot{\varphi}^2 \right) \right] \cos\Upsilon
\nonumber \\
&\quad\quad
-\frac{1}{2} (1+C_{i_{0}}^2) r \dot{r}
\left[2 C_\Theta \dot{\varphi} + (1+C_\Theta^2) \dot{\Upsilon} \right]
\sin 2 \Upsilon
\nonumber \\
&\quad\quad
-\frac{1}{4} (1+C_{i_{0}}^2)
\left[ 4 C_\Theta r^2 \dot{\varphi} \dot{\Upsilon} + (1+C_\Theta^2)
\left( \frac{G M}{r} - \dot{r}^2 + r^2 \dot{\varphi}^2 + r^2 \dot{\Upsilon}^2
\right) \right] \cos 2 \Upsilon
\nonumber \\
&\quad\quad
-\frac{3}{4} S_{i_{0}}^2 S_\Theta^2 \left( \frac{G M}{r} - \dot{r}^2 + r^2
\dot{\varphi}^2 - \frac{1}{3} r^2 \dot{\Upsilon}^2
\right)
\bigg\} \cos 2 \varphi
\nonumber \\
&\quad
- S_{i_{0}} C_{i_{0}} S_\Theta C_\Theta r \dot{r} \dot{\Upsilon} \sin\Upsilon
+ S_{i_{0}} C_{i_{0}} S_\Theta
\left[r^2 \dot{\varphi} \dot{\Upsilon} - C_\Theta \left( \frac{G M}{r} -
\dot{r}^2 - r^2 \dot{\varphi}^2 \right)
\right] \cos\Upsilon
\nonumber \\
&\quad
-\frac{1}{2} (1+C_{i_{0}}^2) S_\Theta^2 r \dot{r} \dot{\Upsilon}
\sin 2 \Upsilon
- \frac{1}{4} (1+C_{i_{0}}^2) S_\Theta^2
\left( \frac{G M}{r} - \dot{r}^2 - r^2 \dot{\varphi}^2 + r^2 \dot{\Upsilon}^2
\right) \cos 2 \Upsilon
\nonumber \\
&\quad
+\frac{1}{4} S_{i_{0}}^2
\left[
4 C_\Theta r^2 \dot{\varphi} \dot{\Upsilon}
+ (1 + C_\Theta^2 ) r^2 \dot{\Upsilon}^2
+ (1-3 C_\Theta^2 ) \left(\frac{G M}{r}-\dot{r}^2 -r^2 \dot{\varphi}^2\right)
\right]
\Bigg)\,,
\\
h_{\times} \big|_{\text{Q}}
&=
\frac{2 G \mu}{c^4 R'}
\Bigg(
\bigg\{
S_{i_{0}} S_\Theta r \dot{r} (2 C_\Theta \dot{\varphi} +
\dot{\Upsilon})\sin\Upsilon 
+S_{i_{0}} S_\Theta \left(
\frac{G M}{r} - \dot{r}^2 + r^2 \dot{\varphi}^2 + C_\Theta r^2
\dot{\varphi} \dot{\Upsilon}
\right)\cos\Upsilon
\nonumber \\
&\quad\quad
- C_{i_{0}} r \dot{r} \left[(1+C_\Theta^2)\dot{\varphi} + 2
C_\Theta \dot{\Upsilon} \right]\sin 2\Upsilon
\nonumber \\
&\quad\quad
- C_{i_{0}} \left[
C_\Theta \left( \frac{G M}{r} - \dot{r}^2 + r^2 \dot{\varphi}^2
+ r^2 \dot{\Upsilon}^2 \right) + ( 1 + C_\Theta^2) r^2 \dot{\varphi}
\dot{\Upsilon}
\right]\cos 2\Upsilon 
\bigg\}\sin 2 \varphi
\nonumber \\
&\quad
+\bigg\{
S_{i_{0}} S_\Theta 
\left[ r^2 \dot{\varphi} \dot{\Upsilon}
+ C_\Theta \left(\frac{G M}{r} - \dot{r}^2 + r^2 \dot{\varphi}^2 \right)
\right] \sin \Upsilon
- S_{i_{0}} S_\Theta r \dot{r} (2 \dot{\varphi} + C_\Theta
\dot{\Upsilon}) \cos\Upsilon 
\nonumber \\
&\quad\quad
- \frac{1}{2} C_{i_{0}} 
\left[
4 C_\Theta r^2 \dot{\varphi} \dot{\Upsilon}
+ (1 + C_\Theta^2) \left(\frac{G M}{r} - \dot{r}^2 + r^2 \dot{\varphi}^2 +
r^2 \dot{\Upsilon}^2  \right)
\right] \sin 2\Upsilon
\nonumber \\
&\quad\quad
+ C_{i_{0}} r \dot{r} \left[ 2 C_\Theta \dot{\varphi}
+ (1 + C_\Theta^2) \dot{\Upsilon} \right] \cos 2\Upsilon
\bigg\}\cos 2 \varphi
\nonumber \\
&\quad
+S_{i_{0}} S_\Theta
\left[
r^2 \dot{\varphi} \dot{\Upsilon} - C_\Theta \left(\frac{G M}{r} - \dot{r}^2 -
r^2 \dot{\varphi}^2 \right)
\right]\sin\Upsilon
+S_{i_{0}} S_\Theta C_\Theta r \dot{r} \dot{\Upsilon} \cos\Upsilon 
\nonumber \\
&\quad
-\frac{1}{2} C_{i_{0}} S_\Theta^2 \left(
\frac{G M}{r} - \dot{r}^2 - r^2 \dot{\varphi}^2 + r^2 \dot{\Upsilon}^2
\right) \sin 2\Upsilon
+ C_{i_{0}} S_\Theta^2 r \dot{r} \dot{\Upsilon}\cos 2\Upsilon
\Bigg)\,.
\end{align}
\end{subequations}
\end{widetext}
We would like to remind (again) the reader that 
the orbital phase is denoted by $\varphi$, the angle $\Upsilon$
describes the phase of the line of nodes $\vek{i}$,
$\dot{\varphi} = d\varphi / dt$, $\dot{\Upsilon} = d\Upsilon / dt$,
and $\dot{r}=(\vek{n} \cdot \vek{v})$.
The parametric PN accurate expressions for these time
derivatives can easily be computed using the parametric solution as
$\dot{r} = dr/dt = (dr/du) (du/dt)$,
$\dot{\varphi} = d\varphi/dt = (d\varphi/dv) (dv/du) (du/dt)$ and
$\dot{\Upsilon} = d\Upsilon/dt = (d\Upsilon/dv) (dv/du) (du/dt)$.

Notice that the temporally evolving (observational) orbital inclination $i$
(see Fig.~\ref{sphere_xyz_pqN}),
defined by $\cos i = \vek{N} \cdot \vek{k}$, does not enter the expressions for
$h_{+}|_{\text Q}$ and $h_{\times}|_{\text Q}$.
However, using $i_0$, $\Theta$ and $\Upsilon$, the time evolution of
$i$ is simply given by
\begin{subequations}
\label{eq:time_evolution_of_i}
\begin{align}
\label{eq:cos_i_parametrized}
\cos i
&= \vek{N} \cdot \vek{k}
= C_{i_{0}} C_\Theta - S_{i_{0}} S_\Theta \cos\Upsilon 
\,,
\\
\label{eq:sin_i_parametrized}
\sin i
&= | \vek{N} \times \vek{k} |
\nonumber \\
&= \left[ ( S_{i_{0}} C_\Theta + C_{i_{0}} S_\Theta \cos\Upsilon )^2
+ S_\Theta^2 \sin^2 \Upsilon \right]^{1/2} 
\nonumber \\
&= \left[ 1 - (C_{i_{0}} C_\Theta - S_{i_{0}} S_\Theta \cos\Upsilon )^2
\right]^{1/2} 
\,.
\end{align}
\end{subequations}
The equation for $\cos i$ is in agreement with 
Eq.~(6) in Ref.~\cite{LBK1995}.
We may also obtain a differential equation for the 
time evolution of $i$, as given in Ref.~\cite{DS88}. 
To get that, we differentiate $\cos i = \vek{N} \cdot \vek{k}$
with respect to time and making use of
$\dot{\vek{k}} = \dot{\vek{L}}/L$ with Eq.~\eqref{eq:dLdt_with_Seff}
and Eq.~\eqref{eq:sin_i_parametrized}. In this way we deduce the following differential equation for the orbital inclination $i$ 
\begin{align}
\frac{d i}{dt} &
= \frac{1}{c^2 r^3} \vek{S}_\text{eff} \cdot \frac{\vek{N} \times
\vek{k}}{\sin i}
= \frac{1}{c^2 r^3} 
\vek{S}_\text{eff} \cdot \frac{\vek{N} \times \vek{k}}{|\vek{N} \times \vek{k}|}
\,,
\end{align}
which agrees with Eq.~(5.15) of Ref.~\cite{DS88}.

The Eqs.~\eqref{eq:h_plus_h_cross_with_SO} are useful 
for scenarios where the angle $\Theta$ is a constant.
This is true for the two cases where our parametric solution is
applicable (see discussions in Sec.~\ref{Sec4} after Eqs.~\eqref{eq:eom_second_and_third_again}).

The approach to compute $h_{+}|_{\text Q}$ and $h_{\times}|_{\text Q}$
may be adapted to include higher order spin effects like spin--spin
interactions. In this case, $\Theta$ will no longer be a constant
angle and expressions for the velocity vector $\vek{v}$ and hence
$h_{+}|_{\text Q}$ and $h_{\times}|_{\text Q}$ will also depend
on $\dot{\Theta}$.

The non--spinning limit, as given by Eqs.~(6) in Ref.~\cite{DGI}, is
obtained in the following way. 
We note that when $S \rightarrow 0$, 
$\vek{J}$ is identical to $\vek{L}$, which implies 
$\Theta \rightarrow 0$, $\theta \rightarrow \pi /2$ and
$i_{0} \rightarrow i$, $i$ being the angle between $\vek{N}$ and $\vek{L}$.
In this limit, Eqs.~\eqref{eq:transfo_angular_coordinates} 
and \eqref{eq:phi_dot} indicate that
$\Upsilon + \varphi = \phi$ and
$\dot{\Upsilon} + \dot{\varphi} = \dot{\phi}$. 
This leads to Eqs.~(6) in Ref.~\cite{DGI} which reads
\begin{subequations}
\begin{align}
h_{+} \big|_{\text{Q}}
&= -\frac{G \mu}{c^4 R'}
\bigg\{
(1+C_{i}^2) \bigg[
2 \dot{r} r \dot{\phi} \sin (2\phi)
\nonumber \\
& \quad 
+\left( \frac{G M}{r} - \dot{r}^2 + r^2 \dot{\phi}^2 \right)
\cos (2\phi)
\bigg]
\nonumber \\
& \quad 
+S_{i}^2 \left( \frac{G M}{r} - \dot{r}^2 - r^2 \dot{\phi}^2 \right)
\bigg\}
\,,
\\
h_{\times} \big|_{\text{Q}}
&= - \frac{2 G \mu C_{i}}{c^4 R'}
\bigg[ 
\left(\frac{G M}{r}-\dot{r}^2 +r^2 \dot{\phi}^2 \right) \sin (2 \phi) 
\nonumber \\
& \quad 
- 2 \dot{r} r \dot{\phi} \cos(2 \phi)
\bigg]
\,,
\end{align}
\end{subequations}
where $C_{i}$ and $S_{i}$ are shorthand notations for $\cos i$ and
$\sin i$. 

The temporal evolution of $h_{+}$ and $h_\times$,
given by Eqs.~\eqref{eq:h_plus_h_cross_with_SO},
may be obtained by adapting the `phasing formalism',
presented in Ref.~\cite{DGI}, and this
will be reported soon \cite{KG2004phasing}.
However, it should be noted that the parametrization, given by
Eqs.~\eqref{eq:solution_for_r_3PN_plus_SO}--\eqref{eq:theta_in_terms_of_SLJ_again},
will be sufficient to determine the 3PN accurate conservative time evolution,
which includes the leading order spin--orbit interaction.

\section{Conclusions}
\label{Sec7}

We have presented Keplerian--type parametrization for the solution of
post--Newtonian accurate conservative dynamics of spinning compact
binaries moving in eccentric orbits.
The above PN accurate dynamics consisted of
3PN accurate conservative orbital dynamics,
associated with non--spinning compact objects,
influenced by the leading order spin--orbit interactions.
The orbital elements of the representation were
explicitly obtained in terms of the
conserved orbital energy, angular momentum and a quantity that
characterizes the leading order spin--orbit interactions
in ADM--type coordinates. 
Our parametric solution is applicable in the following two 
distinct cases, namely, case (i), the equal--mass--case and case (ii),
the single--spin--case. We also derived expressions for 
gravitational wave polarizations suitable to describe
gravitational radiation from spinning compact binaries
moving in eccentric orbits.

The present work has several possible applications and some of them are
currently under investigation.
Employing the `phasing formalism', given by Ref.~\cite{DGI}, along with
our parametric solutions and expressions for 
$h_{+}$ and $h_\times$, we will, in the near future, obtain 
gravitational wave polarizations suitable to describe
gravitational radiation from spinning compact binaries
moving in inspiralling eccentric orbits \cite{KG2004phasing}.
Our parametric solution for the two cases considered implies that
the associated PN accurate conservative binary dynamics
will not be chaotic. This is so as our solution can analytically 
determine the associated dynamics.
It will be interesting to explore (again) numerically the scenarios,
where our parametrization is valid, and investigate
if the numerical solutions still predict chaos.
Finally, as mentioned earlier, the fully 2PN accurate `timing formula'
may be derived using our parametric solution as a 
crucial input. This may be required as spin--orbit interactions 
essentially enter at 2PN order for binary pulsars and 
their effects leave observational signature
\cite{DS88,Damour_Ruffini_1974,OConnel_2004,Stairs_Thorsett_Arzoumanian_2004}.

It is highly desirable to extend the present work in the following directions.
In the literature, there exists a PN accurate orbital equations of motion
for spinning compact binaries, where spin--orbit interactions are
also PN accurate~\cite{TOO04}. It will be interesting to see if a
parametric solution is possible for the above dynamics.
Naturally, it will be interesting to include spin--spin effects
in to our parametric solution.
However, all these extensions are not straightforward
and the parametric solutions for these cases, most probably, will not
be as elegant as the one presented in this paper.

\begin{acknowledgments}
We are indebted to Gerhard Sch\"afer for helpful discussions and
encouragements. We gratefully acknowledge the financial support of
the Deutsche Forschungsgemeinschaft (DFG) through SFB/TR7
``Gravitationswellenastronomie''.

The algebraic computations, appearing in this paper, were performed
using \textsc{Maple} and \textsc{Mathematica}.
\end{acknowledgments}


\bibliographystyle{apsrev}

\begin{thebibliography}{999}
\bibitem{ADM_type}
It turned out that the 3PN accurate Hamiltonian, when it was derived for the
first time, involved derivatives of the phase space variables.
However, using PN accurate higher order contact transformation, the above
Hamiltonian was later reduced to an ordinary Hamiltonian, which depends
only on the phase space variables.
We refer the above ordinary Hamiltonian as the 3PN accurate 
Hamiltonian in ADM--type coordinates.
\end{thebibliography}

\end{document}